\documentclass[a4paper]{cas-dc}

\usepackage[authoryear]{natbib}
\def\tsc#1{\csdef{#1}{\textsc{\lowercase{#1}}\xspace}}
\tsc{WGM}
\tsc{QE}
\tsc{EP}
\tsc{PMS}
\tsc{BEC}
\tsc{DE}
\begin{document}

\shorttitle{}

\shortauthors{Xie et~al.}

\title [mode = title]{Nonreciprocal interactions in crowd dynamics: investigating the impact of moving threats on pedestrian speed preferences}                      

\author[1,2]{Shaocong Xie}
\author[1,2]{Rui Ye}
\author[3]{Xiaolian Li}
\author[1,2]{Zhongyi Huang}
\author[4]{Shuchao Cao}
\author[5]{Wei Lv}
\author[1,2]{Hong He}
\author[1,2]{Ping Zhang}
\author[1,2]{Zhiming Fang} \ead{zhmfang2015@163.com} \cormark[1]
\author[6]{Jun Zhang} \ead{junz@ustc.edu.cn} \cormark[1]
\author[6]{Weiguo Song} \ead{wgsong@ustc.edu.cn} \cormark[1]

\affiliation[1]{organization={Business School, University of Shanghai for Science and Technology},
    city={Shanghai},
    postcode={200093}, 
    country={China}}
\affiliation[2]{organization={School of Intelligent Emergency Management, University of Shanghai for Science and Technology},
    city={Shanghai},
    postcode={200093}, 
    country={China}}
\affiliation[3]{organization={Public Security Department, Fujian Police College},
    city={Fuzhou},
    postcode={350007}, 
    country={China}}
\affiliation[4]{organization={School of Automotive and Traffic Engineering, Jiangsu University},
    city={Zhenjiang},
    postcode={212013}, 
    country={China}}
\affiliation[5]{organization={School of Safety Science and Emergency Management, Wuhan University of Technology},
    city={Wuhan},
    postcode={430070}, 
    country={China}}
\affiliation[6]{organization={State Key Laboratory of Fire Science, University of Science and Technology of China},
    city={Hefei},
    postcode={230027}, 
    country={China}}

\cortext[cor1]{Corresponding author}

\begin{abstract}
Nonreciprocal interaction crowd systems, such as human-human, human-vehicle, and human-robot systems, often have serious impacts on pedestrian safety and social order. A more comprehensive understanding of these systems is needed to optimize system stability and efficiency. Despite the importance of these interactions, empirical research in this area remains limited. Thus, in our study we explore this underresearched area, focusing on scenarios where nonreciprocity plays a critical role, such as mass stabbings, which pose a substantial risk to public safety. We conducted the first experiments on this system and analysed high-accuracy data obtained from these experiments. Specifically, we conduct laboratory experiments on three scenarios: single exits, dual exits, and obscured chases. Then, we introduce an equation to describe the mechanism behind a direct threat zone, which significantly affects pedestrian behaviour. The extent of the direct threat zone is determined by the speed of the moving threat and the radius of danger occurrence. The equation can provide insights for different nonreciprocal interaction crowd systems, where the subjects of the nonreciprocal interactions can be a pedestrian, an autonomous robot or a vehicle. We further categorize potential threats into direct, adjacent, and rear-view zones, quantifying the level of threat for pedestrians. Our study revealed that a pedestrian's desired velocity correlated positively with potential threat intensity, increasing until near the direct threat zone. An emerging steady state is observed when escape routes are blocked by moving threats. This deviation affects the density-velocity relationship, making it distinct from the general relationship. This deviation signifies unique pedestrian behaviour in the presence of moving threats. Additionally, the rate of change in the angle for pedestrian motion in various desired directions is synchronized. This indicates the emergence of collective intelligence in nonreciprocal interaction crowd systems. As a result, our study may constitute a pioneering step towards understanding nonreciprocal interactions in crowd systems through laboratory experiments. These findings may enhance pedestrian safety and inform not only government crowd management strategies but also individual self-protection measures.
\end{abstract}

% \includegraphics{figs/grabs.pdf}

\begin{highlights}
\item This study includes the first complete and high-data-accuracy laboratory experiment on strongly coupled nonreciprocal crowd systems.
\item An equation is used to describe the major motion pattern between pedestrians and strongly coupled nonreciprocal interaction subjects. This equation can provide insights for modelling human-human, human-vehicle, and human-robot strongly coupled nonreciprocal interactions.
\item The empirical study confirms that the high desired velocity of pedestrians is due to the presence of moving threats and reveals the mechanism of pedestrian-desired velocity change
\item The self-organization phenomenon of the emergence of collective intelligence is observed in the experiments, and the reasons for the emergence of this self-organization phenomenon are also provided.
\end{highlights}

% Keywords

\maketitle
\section{Introduction}

Urban populations made up 56\% of the world's total population in 2021, and that number is expected to increase to 68\% by 2050 \citep{worldcities2022}. Rapid urbanization highlights the pressing need for effective pedestrian traffic management. This involves ensuring not only pedestrian mobility but also pedestrian safety. As cities continue to expand, understanding how to manage increasingly complex pedestrian systems becomes a pressing issue. At the heart of crowds in urban areas is an often-overlooked component, namely, nonreciprocal interactions \citep{Kanoaminimal}. These interactions occur when individuals or entities within a crowd respond asymmetrically to each other's actions, leading to complex behaviour patterns \citep{calozElectromagneticNonreciprocity2018,fruchartNonreciprocalPhaseTransitions2021,kryuchkovDissipativeSpinodalDecomposition2018a}. While these nonreciprocal interactions are common and have serious safety impacts, few researchers have focused on such interactions in crowds. Such safety impacts may range from moving vehicles in crowds \citep{gorriniobservation2018} or robots in crowds to mass protests and other unpredictable hazards \citep{bernardiniterrorist2021,lueffects2021, lumulti-agent2021, niustrategy2021}. Nonreciprocal interactions affect crowd behaviour at short time intervals, making them difficult to analyse consistently. Such instances are particularly evident during interactions between pedestrians and vehicles, where the nonreciprocal nature is pronounced yet fleeting as it intersects with a pedestrian's intended path \citep{golchoubianPedestrianTrajectoryPrediction2023,markkulaExplainingHumanInteractions2023,gorriniobservation2018}. The same situation occurs in crowds. We can effectively observe the action behaviour of nonreciprocal interactions at the moment of pedestrian-overtaking behaviour, but this period is particularly transient. However, we can observe significant and persistent nonreciprocal interactions in pedestrian experiments. Thus, conducting a pedestrian experiment is important for studying pedestrian dynamics.

In past years, the physical mechanisms underlying destabilization processes in crowd systems have been investigated by using realistic data in several studies\citep{helbingsimulating2000, helbingself-organized2005, helbingdynamics2007}. These studies have yielded valuable insights.
However, real-world data encompass a variety of crowd behaviours, and it is difficult to disentangle these behaviours to study the mechanisms behind specific behaviours individually. Thus, experimental studies have been conducted to analyse the behaviour of pedestrians in different situations, such as bottleneck areas and corridor flows \citep{daamenExperimentalResearchPedestrian2003,hoogendoornpedestrian2004,kretzexperimental2006,kretzexperimental2006-1,garcimartinflow2016,renflows2021}. Observations made during evacuation experiments revealed the existence of intermittent flow patterns and the effect of door width on pedestrian dynamics \citep{moussaidtraffic2012}. Fundamental diagrams representing the relationship between pedestrian density and velocity have been extensively studied, providing insights into the collective behaviours of pedestrians in different environments \citep{seyfriedfundamental2005,zhangtransitions2011,caofundamental2017}.
Moreover, researchers have explored the decision-making process, route-choice behaviour, and activity scheduling of pedestrians\citep{hoogendoornpedestrian2004,daamenCapacityDoorsEvacuation2010}. In past research, the foundation of our understanding of crowd systems was built based on an experimental approach. However, these experiments were not designed to explore nonreciprocal interactions in crowds. Most of the crowd nonreciprocal interactions in the experiments were weakly coupled and had short durations; thus, they were not sufficient to effectively extract nonreciprocal interactions to understand nonreciprocal interactions in crowded systems.

Several studies of strongly coupled nonreciprocal interactions in crowds have also been conducted by using real-world data. \citep{bernardiniterrorist2021} collected videotapes of recent terrorist attacks in Europe and conducted qualitative and quantitative analyses to identify common and distinct pedestrian behaviours compared to those during other types of emergencies. \cite{wangEmpiricalStudyCrowd2019} analysed a recorded video of the March 2014 Kunming terrorist attack in China and observed phenomena such as crowd oscillation and self-organization grouping. In a laboratory experimental study, \citep{dingExperimentSimulationStudy2021} conducted evacuation experiments under armed assault attacks in a classroom setting. However, these studies lacked complete and accurate pedestrian trajectories, thereby making it nearly impossible to further investigate the dynamics of nonreciprocal interaction crowd systems. \citep{parisipedestrian2021} explored the fundamental characteristics of the nonreciprocal interactions between crowds and bulls at the Running of the Bulls Festival in Pamplona, Spain. This is a valuable study involving strongly coupled pedestrian nonreciprocal interactions and high-accuracy data. However, due to practical limitations, it is difficult to capture the long-term evolution of the bull-crowd nonreciprocal interaction system.

To address these limitations, we conducted pioneering experiments on strongly coupled nonreciprocal interactions. These experiments capture the full evolutionary process of the nonreciprocal interactions in the crowd system with highly accurate trajectory data. In our study, we delve into the dynamics of crowd behaviour when individuals interact with moving threats through strongly coupled nonreciprocal interactions, an important but previously underexplored domain. We reveal how this interaction affects the pedestrian's desired speed and the pedestrian's desired movement goal. Our findings indicate that pedestrian density and velocity relationships are distinctively modified by moving threats, differing from conventional crowd systems. Furthermore, we derive equations to interpret the influence of moving threats on pedestrian behaviour through strongly coupled nonreciprocal interactions. These mathematical representations elucidate how crowd behaviour is affected by such interactions in environments where moving threats exist. Additionally, these equations provide insights into the modelling of strongly coupled non-reciprocal interactions between pedestrians and autonomous robots or vehicles. Specifically, when there is a potential collision with the pedestrian's intended path. In such cases, a moving robot or vehicle can be considered as a moving threat. We identify the emergence of a self-organization phenomenon within the system and investigate its characteristics and the necessary conditions for its emergence. This analysis enhances our understanding of collective intelligence in such systems.

In this study, we deduce fundamental diagrams from the experimental data. Statistical methodologies are utilized to determine the variance in patterns from the conventional crowd system. Subsequently, we formulate and validate an equation (direct threat zone) to enhance our comprehension of crowd behaviour influenced by strongly coupled nonreciprocal interactions. Additionally, we model the distribution of assault targets across relative positions. Building upon this, we quantify the potential threat degree to pedestrians and examine the dynamics that govern the fluctuation in pedestrians' desired speed. Moreover, we analyse the self-organization phenomenon that emerges in this system by examining the strength of the coupling relationship between pedestrians' two desired moving targets. The phenomenon of self-organization pertains to pedestrians collectively adjusting their direction or their speed to establish an arching pattern as a strategy to evade moving threats.

This paper is organized as follows. In Section \ref{sec2}, we present our experiment on the nonreciprocal population system, including the experimental design, experimental data, and data collection methods. In Section \ref{sec3}, we discuss the dynamics of the nonreciprocal interaction crowd system and the impact of human movement threats on pedestrian behaviour. In Subsection \ref{sec31}, we analyse the fundamental diagrams for scenarios both with and without attackers. We suggest that the difference in the pedestrian's desired speed may be related to the degree to which the pedestrian is affected. In Subsection \ref{sec32}, we analyse the behaviour of pedestrians facing mobile threats and propose a formula for the direct threat zone. We find that pedestrians are most affected within the direct threat zone. In Subsection \ref{sec33}, we further analyse the other zones. Based on the three threat zones, an equation is proposed. In Subsection \ref{sec34}, we analyse the self-organization phenomenon (collective intelligence) that emerges during the experiment. The reason for the disappearance of the desired evasion velocity of pedestrians in this phenomenon is analysed. In Section \ref{sec4}, we interpret the findings in the context of the literature and discuss both the limitations of the study and the implications of the findings for pedestrian dynamics and traffic management. In Section \ref{sec5}, we summarize the key findings of our study and provide suggestions for future research in this field. In Section \ref{sec6}, we provide a detailed description of the methods used in this paper.

\section{Data collection experiments\label{sec2}}

\begin{figure*}
    \centering
        \includegraphics[width=1\textwidth]{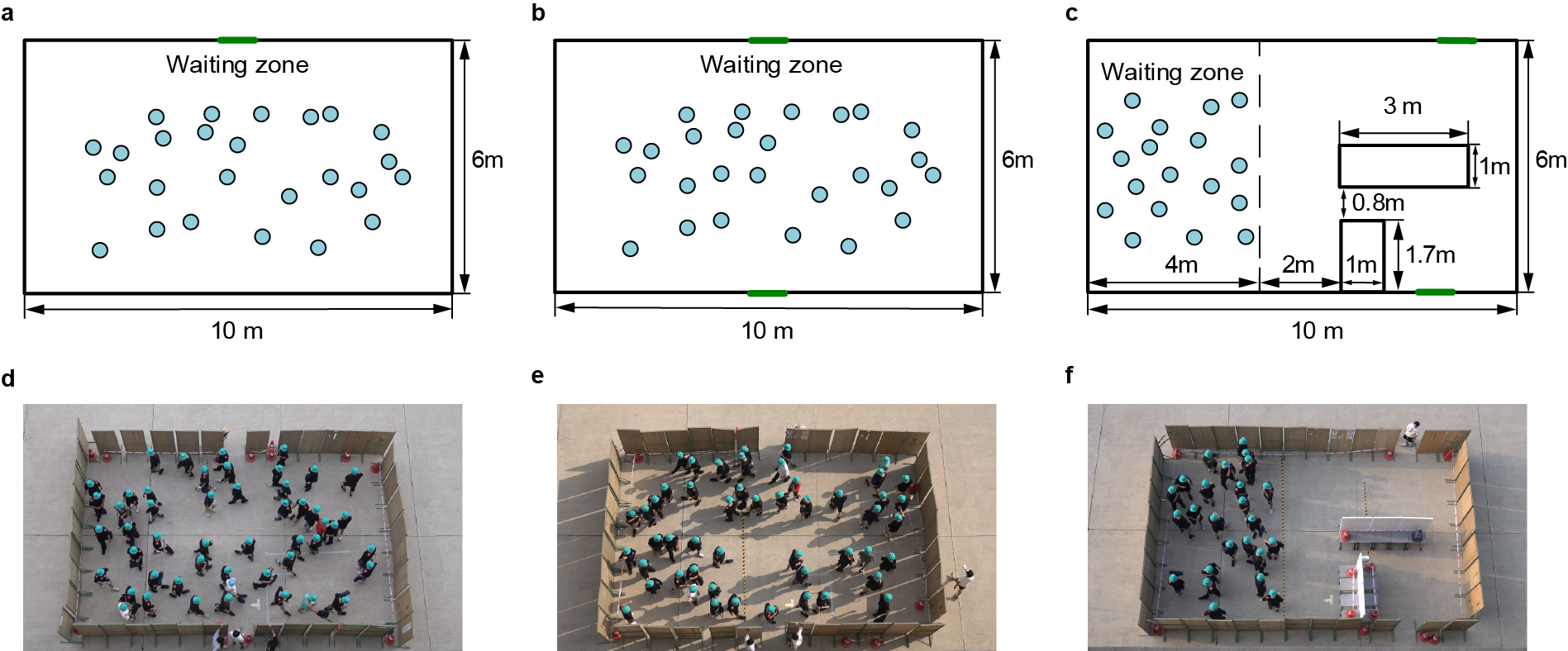}
\caption{\textbf{Experimental design.} \textbf{(a-c)}, Design scenarios for single-exit, dual-exit, and exit-obscured chase experiments. The circles represent participants, the black lines represent obstacles, and the green lines represent exits. A waiting zone was set up for participants to await the start of each experiment. Initially, participants were requested to walk randomly within the zone. \textbf{a}, The single-exit experiment involved setting only one exit on the fence, which was aligned with the centre of the fence. \textbf{b}, In the dual-exit experiment, two exits were established at opposite locations. \textbf{c}, The exit-obscured chase experiment incorporated two barriers, delineating a $4-\text{m} \times 6-\text{m}$ rectangular waiting zone positioned away from the exits. The information in the front area of the lower exit was obscured by a 2-m-high plate. Therefore, this phenomenon could not be observed in the waiting zone. Information in the front area of both exits could not be observed from another exit. \textbf{(d-f)}, Video screenshots of single-exit, dual-exit, and exit-obscured chase experiments.}\label{Fig.1}
\end{figure*}

The experiment simulated a realistic scenario of mass stabbing. We study nonreciprocal crowd systems while also providing suggestions for crowd safety in real-life scenarios. In the experiment, strongly coupled nonreciprocal interactions between attackers and evacuees occurred in the system. The attacker's goal was to close in on the evacuees, while the evacuees aimed to increase their distance from the attacker. The attacker and the evacuee had completely opposite interaction rules, which could be considered strongly coupled nonreciprocal interactions. Importantly, not all nonreciprocal interactions within a crowd are threat related; however, interactions involving threats are a salient example of such dynamics.

\subsection{Experimental Setup and Design}
To investigate these dynamics, we conducted three nonreciprocal interaction crowd dynamics experiments: single-exit, dual-exit, and exit-obscured chase experiments (Fig. \ref{Fig.1}). The experimental scenario involved a $10-\text{m} \times 6-\text{m}$ rectangular room, and the surrounding walls were fenced by baffles. The experiment simulated a mass stabbing scenario with different numbers of attackers in various scenarios (Appendix, Table \ref{tablea1}).

Participants were divided into two groups, namely, attackers and evacuees, based on the presence of a "weapon" (stick) concealed in their backpacks. Those with weapons assumed the role of attackers, while those without weapons were considered evacuees.

At the onset, each participant was equipped with both a hat and a backpack. The purpose of the backpacks was to conceal the identity of the attacker among the participants. Each backpack was identical in appearance to ensure that the stick it might contain was not visible, thus preventing the participants from preemptively identifying and avoiding the attacker. This design was chosen to emulate a real-life situation where an attacker's intentions and weapon remain concealed until an attack is underway. To reflect the reality that not everyone carries a backpack, we set the backpack wearing ratio at 50\% in the single-exit and dual-exit scenarios.

In both the single-exit and dual-exit scenarios, each exit was standardized to 0.8 m in width, allowing two pedestrians to pass through simultaneously. In the exit-obscured chase scenario, the room featured obstacles that created a passage 0.8 m in width, with one side door being wider at 1.2 m, meaning that it would accommodate three pedestrians. This setup was intended to introduce an element of uncertainty and stress, akin to real-world conditions where obscured exits and sudden threats can lead to confusion and elevated stress levels. In addition, confusion and stress do not mean that there will be a "panic". That is because people's behaviour in emergencies often does not meet the traditional stereotype of irrationality or panic, but is more complex, organized, and purposeful \citep{druryEveryoneThemselvesComparative2009,druryPsychologicalDisasterMyths2013,dezecacheNatureDeterminantsSocial2021}. Nevertheless, the size of the study area may have made the presence of an exit obvious. Thus, the height of the obstacles utilized ensured that participants could not obtain the exit-opening status in the initial area. We adjusted the opening status of the two exits in each experiment and determined the conditions for the different exit opening statuses in a random order.

Simulating life-threatening behaviour in a controlled experiment has inherent limitations. Real-life threat scenarios invoke intense psychological and physiological responses that are impossible to fully replicate in a laboratory setting due to ethical and practical constraints. Participants are aware at some level that they are in a safe, controlled environment, which likely influences their responses and behaviours, potentially reducing the validity of our findings. To mitigate these limitations, our study design aimed to create a setting that was as realistic as possible while ensuring the safety of the participants. The design contained elements of stress and uncertainty. This included towering enclosures, randomly appearing attackers and a randomly arranged sequence of experiments (with varying numbers of attackers). In particular, we conducted a pre-experiment with no enclosures or enclosures. Interviews were conducted with the experimenters who participated in the preexperiment. Most of the participants noted that the presence of enclosures raised their sense of stress.

\subsection{Participant Recruitment}
These experiments were carried out at a college located in Fuzhou in 2022. During each round of the experiments, the single-exit and dual-exit experiments involved 50 participants each, while the exit-obscure chase experiment involved 25 participants.

Our experimental sample sizes were determined by density. We tried to set up densities that were close to reality. This meant that the density needed to range from 0.6 $\text{m}^{-2}$ to 0.85 $\text{m}^{-2}$. An initial pilot study was undertaken with 50 participants in each scenario to determine the feasibility of the experiments. The many obstructions in the exit-obscured cavity hindered the motion of the participants. Thus, for safety reasons, we reduced the number of participants to 25 in this scenario. In these experiments, participants were recruited from a local university and volunteered to participate. The ages ranged from 20 to 22 years, and the male-to-female ratio was approximately 4:1. Since we recruited participants at universities, the participants were all college students. The university oversaw verifying the physical health and mental health of the students. Thus, it was assumed that each participant was physically and mentally fit. We did not specifically screen students for prior experience. If someone expressed discomfort or fear, we allowed them to withdraw from the study.

\subsection{Experimental Procedure and Instructions}

Initially, each participant entered the experimental field sequentially and received a backpack, which confirmed the team to which they were assigned. They were then instructed to wait for everyone to enter and then walk randomly within the waiting zone to wait for the start of the experiment.

Upon receiving the "ready" order, the attacker surreptitiously chose their initial position and initiated the first assault within a ten-second window, marking the start of the experiment.

Evacuees were asked to begin evacuation only after they saw the "weapon"; otherwise, they were all asked to walk randomly within the area. For each round of the experiment without an attacker, the "begin evacuation" order was announced after 10 seconds. At that time, the experiment focused on pedestrian evacuation.

An evacuee who was hit more than once was considered a casualty and was required to remain in the position in which he or she was hit. A round was considered to have concluded once all non-causality participants had exited the experimental field.

We gave our orders in public over a loudspeaker. The order "attacker ready" was used in every round of the experiment, regardless of the existence of an attacker in that round. A 10-second waiting period occurred because the attacker was required to launch each attack within ten seconds. This information was available to everyone. Therefore, pedestrians also needed to wait for ten seconds in experiments without attackers.

This approach was used to avoid pedestrians' expectations of the number of attackers in each experiment having an impact on the results of experiments that included attackers.
The abovementioned processes were used to reproduce as many real-life situations as possible in which the attacker picks an advantageous initial position when launching an initial assault. Thus, the evacuees were not aware of the attacker's position before the assault began. This approach prevented evacuees from having expectations about the number and location of attackers, which could have interfered with the experimental results.

The following instructions were provided to the participants:
\begin{enumerate}
\item Attackers aim to "immobilize" evacuees by hitting
them with a "weapon" (a harmless prop stick).
\item Attackers must initiate their assault within ten seconds
following the "attacker ready" announcement.
\item Evacuees should attempt to leave the area without
being hit and can only start evacuating after seeing the
"weapon".
\item Evacuees who are hit by the "weapon" more than once are considered "immobilized" and must remain in the spot where they were hit.
\item In rounds without an attacker, evacuees must wait for the
"start evacuation" command before moving towards the
exits.
\item Participants must confirm their group assignment
upon receiving their backpack and must not share this
information.
\item After receiving a backpack, all participants were asked to move randomly into the initial area and wait for the start of the experiment.
\end{enumerate}

The rules of behaviour were as follows:
\begin{enumerate}
\item Participants were required to follow the instructions provided prior to the experiments.
\item All participants were obliged to ensure their safety and the safety of others throughout the experiment.
\item If a participant feels uncomfortable at any point, they are allowed to withdraw from the experiment.
\end{enumerate}

To familiarize participants with the procedures, three practice rounds were conducted before the actual experiment.

In our experiment, "immobilized" refers to the assumption that participants have lost the ability to move due to being assaulted by the attacker. The requirement for evacuees to be hit "more than once" to be classified as casualties served to ensure that every immobilized participant was an assault target. This condition excluded any accidental or incidental contact, thereby confirming that each hit was a deliberate act by the attacker. The "weapon" device used was a 60-cm stick made of plastic and encased in soft polyurethane foam rubber to prevent injury during the simulation.

\subsection{Data Collection Method}
To record the experimental observations, an AX700 camera
with a resolution of 3840 x 2160 pixels and a frame rate of 25 frames/second were used. Participants were instructed to wear coloured hats, and video recordings were taken from the top floor of the building to ensure optimal tracking. Afterwards, the video clips were imported into PeTrack tracking software \citep{boltesautomatic2010,boltescollecting2013} to track the participants' heads and collect movement trajectory data. This software can track people on flat or uneven terrain either with or without markers. The reliability and accuracy of the trajectories extracted by the software have been previously validated in numerous studies of pedestrian dynamics. The exact trajectory of each person was collected, and a detailed analysis of their dynamics was performed. Following the guidelines provided in the PeTrack software, two calibrations were performed to extract precise trajectories from the recordings. This process ensured that the pedestrian trajectories obtained via the software were accurate. Manual modification was also conducted to increase the accuracy of the positioning data for each pedestrian.

\subsection{Safety Measures and Ethical Considerations}
To ensure participant safety, the "weapon" used in the experiments was a prop that was completely harmless. It was wrapped in thick, soft foam to ensure that it would not cause any harm to the participants even in the case of contact. A specialized medical team was always present and was advised to respond immediately in the event of an emergency or if a participant experienced any discomfort. Prior to the experiments, participants received a thorough briefing about the nature of the study and were told that they could withdraw at any time should they feel uncomfortable. Furthermore, all the experimental procedures were approved by the University of Shanghai for Science and Technology Research Ethics Committee, and all the methods were performed in accordance with the relevant guidelines and regulations.
Written informed consent was obtained from each participant. In addition, all participants in this study were fully informed that they would be videotaped for data collection. It was noted that videotaping did not capture the participants' facial features. All the data were anonymized and used strictly for research purposes. The videos were stored securely and accessible only to the research team. In addition, any identifiable features were blurred or removed during the analysis process to maintain the anonymity of the participants. We strictly adhered to ethical guidelines for research involving human subjects, and the study was approved by our institutional ethics committee.

\subsection{Capture Metrics}
\begin{table*}[h]
\caption{\textbf{Description of casualty data.} This table describes the casualty data, specifically, the time it took for an attacker to capture a person and the ratio of casualties caused under different experimental setups. Caught Time is the time it took the attacker to get from one attack to the next, and Casualty Ratio is the number of casualties divided by the number of evacuees involved in the corresponding experiment.}
\label{table_exp}
\begin{tabular}{ccccc}
\toprule
Experimental   scenes & Number of Attackers & Max Caught Time (s) & Mean Caught Time (s) & Casualty Ratio \\
\midrule
Single   Exit         & 1 & 2.68 & 0.68 & 0.26 \\
Single   Exit         & 2 & 3.76 & 1.03 & 0.40 \\
Single   Exit         & 3 & 1.16 & 0.52 & 0.53 \\
Dual   Exit           & 1 & 1.48 & 0.55 & 0.19 \\
Dual   Exit           & 2 & 2.68 & 0.81 & 0.27 \\
Dual   Exit           & 3 & 2.04 & 0.62 & 0.27 \\
Exit-Obscured   Chase & 1 & 1.44 & 0.76 & 0.08 \\
Exit-Obscured   Chase & 2 & 1.92 & 0.68 & 0.35 \\
Exit-Obscured   Chase & 3 & 2.16 & 0.61 & 0.47 \\
Exit-Obscured   Chase & 5 & 1.80 & 0.62 & 0.59 \\     
\bottomrule
\end{tabular}
\end{table*}

In the context of an experimental investigation of simulated mass stabbing, our study presents an analysis of experimental data, focusing on the correlation between the number of attackers, the time between attacks, and the casualty ratio. The experiment was conducted in three distinct scenarios, namely, single-exit, dual-exit, and exit-obscured chase scenarios. Generally, the data collected from realistic experiments contains a portion of noisy data. To reflect the true statistical description, we removed outliers from the statistical data by the interquartile range (IQR) method.

In these scenarios, an increase in the number of attackers resulted in a greater casualty ratio. The casualty ratio increased from 0.26 to 0.53, indicating a greater risk for pedestrians in scenarios with more attackers.

In the single-exit and dual-exit scenarios, the maximum time between attacks decreased and then increased as the number of attackers increased. The mean capture time between attacks followed a similar trend. However, the casualty ratio increased from 0.19 to 0.27 when the number of attackers increased from one to two but remained constant at 0.27 with three attackers.

The exit-obscured chase scenario presented an entirely different dynamic. The maximum time between attacks increased initially from 1.44 seconds with one attacker to 1.92 seconds with two attackers and then decreased to 1.80 seconds when five attackers were present. The mean time between attacks steadily decreased as the number of attackers increased, while the casualty ratio consistently increased, reaching a peak of 0.59 with five attackers.

In both the single-exit and dual-exit scenarios, both the maximum capture time and the mean capture time increased when the number of attackers reached two. In contrast, both values were steadier
in the exit-obscured chase scenario. We can assume that the attackers' attack coverage areas were nearly identical in all the scenarios. Thus, the increasing and decreasing trends may be due to the presence of obstacles in the exit-obscured chase experimental scenario, in which the space was divided into multiple subspaces. This arrangement made it easier for the attacker to cover most of the subspace on his or her own rather than moving around. With more than 2 attackers, the space can be better covered by the combined attack range. In summary, the observed trends in the mean capture time and maximum capture time may be due to differences in the layout of the experimental space and the coordination of attack strategies for multiple attackers.

\section{Results\label{sec3}}

\subsection{Empirical evidence that a pedestrian is most likely to achieve a higher desired velocity when confronted with moving threats\label{sec31}}

Control conditions for moving threats were included in the experiments. As in most evacuation experiments, pedestrians evacuated the room as quickly as possible by walking towards the exit. Trajectories were collected from experiments involving moving threats and those without moving threats. Both were used to calculate pedestrian density and velocity in different frames. The speed-density fundamental diagram was calculated by the density and velocity calculation method (see Methods for details). Since pedestrians were instructed to stop running after being hit twice or more, a pedestrian's trajectory was recorded only before he or she was hit.

\begin{figure*}
    \centering
        \includegraphics[width=1\textwidth]{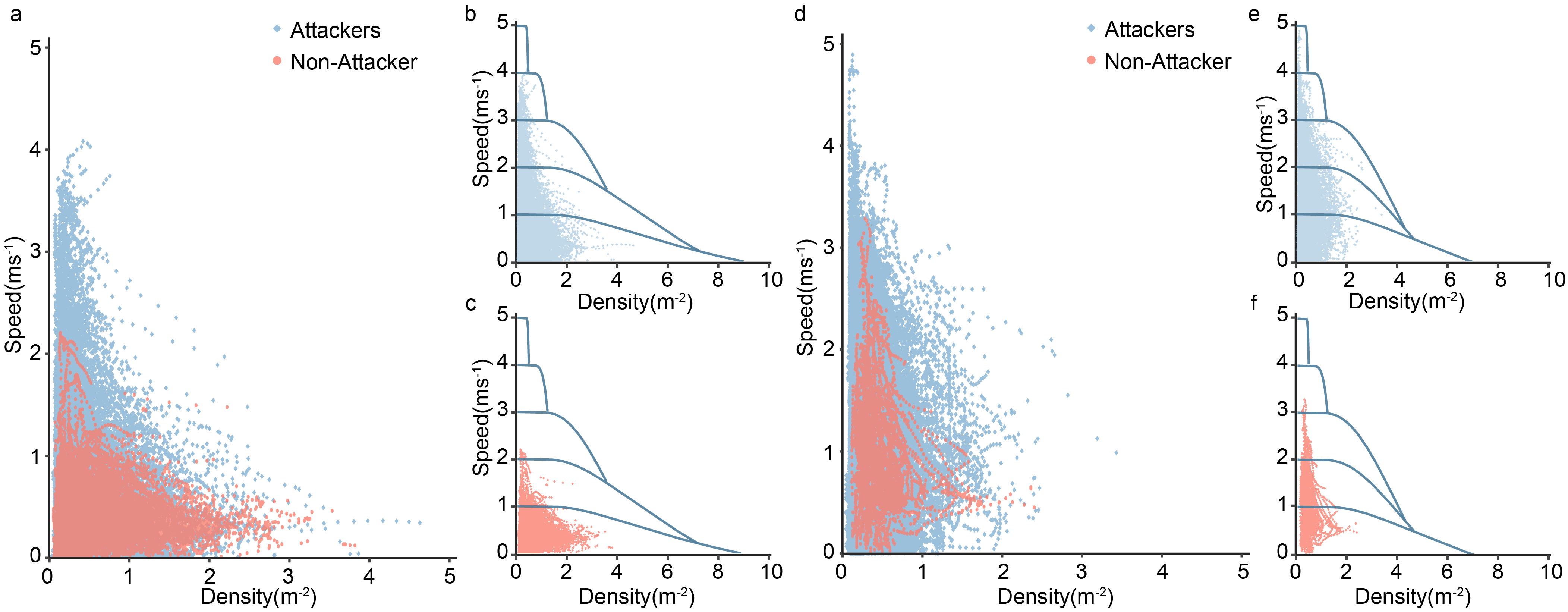}
\caption{\textbf{Speed-density fundamental diagrams for the presence and absence of attackers.} \textbf{(a and d)} Scatterplots illustrating the speed-density relationships in the single-exit and exit-obscured chase experiments, respectively, with blue areas denoting setups including attackers and red areas denoting setups without attackers. The detailed velocity and density calculation equations used are provided in the Methods section. Differences between the presence and absence of attackers were also assessed in both setups. \textbf{(b, c, e and f)} The desired velocity curves display varying speeds at different densities, with specific curve parameters outlined in the Appendix, Table \ref{tablea1}. The speed of each curve at zero density corresponds to its desired velocity. \textbf{(b-c)} The desired velocity curves in the single-exit experiment share the same $g$ and $\rho_{max}$ parameters, and each curve has a distinct desired velocity. The desired velocity of each curve can be observed at a density equal to 0. Parameters \(g \) and \( \rho_{max} \) are crucial components of the desired velocity curves, as illustrated in Eq. (5). In particular, the achievable desired velocity for pedestrians in setups without attackers is lower than that for pedestrians in setups with attackers. \textbf{(e-f)} The desired velocity curves in the exit-obscure chase experiment vary from those in the single-exit experiment but maintain consistent $g$ and desired velocity parameters with those in the single-exit experiment. Similarly, pedestrians in setups without attackers achieve lower desired velocities than those in setups with attackers.}\label{Fig.2}
\end{figure*}

According to the speed-density fundamental diagram comparison, the chase experiment showed greater mean speed and a distinct distribution than did the single-exit experiments, consistent with previous findings \citep{renflows2021}. In the exit-obscured chase experiments, the number of participants was half that used in the single- and dual-exit experiments, and a different waiting area was used. Consequently, the density of the entire area was lower, and pedestrian movement from the wait to the exit was less impeded. However, differences might exist between the speed-density distributions of experiments involving moving threats and those without moving threats.

Statistically significant differences in the speed distributions for single-exit experiments were confirmed by the Kolmogorov-Smirnov test ($D_{(52777)}$=0.277; two-tailed P$<$0.001; 95\% CI=0.052 to 0.065; Cohen's d=0.156), Student's t test ($t_{(52776)}$=17.882; two-tailed P$<$0.001; 95\% CI=0.052 to 0.065; Cohen's d=0.156) and Welch's t test ($t_{(50206)}$=29.395; two-tailed P$<$0.001; 95\% CI=0.057 to 0.065; Cohen's d=0.143). Similarly, for the exit-obscured chase experiments, the Kolmogorov-Smirnov test ($D_{(28267)}$=0.260; two-tailed P$<$0.001; 95\% CI=0.089 to 0.125; Cohen's d=0.138), Student's t test ($t_{(28266)}$=11.586; two-tailed P$<$0.001; 95\% CI=0.089 to 0.125; Cohen's d=0.138) and Welch's t test ($t_{(25135)}$=15.966; two-tailed P$<$0.001; 95\% CI=0.086 to 0.110; Cohen's d=0.114) were used. Similar observations have been previously reported in real-life violence videos for speed and density distributions \citep{bernardiniterrorist2021}.

\begin{table*}[h]
\caption{\textbf{Statistical test results.}}
\centering
\begin{tabular}{cccccc}
\hline
Test & Statistic & P-value & 95\% CI & Cohen's d & Experiment\\
\hline
Kolmogorov-Smirnov & $D_{(52777)}$=0.277 & P$<$0.001 & 0.052 to 0.065 & 0.156 & Single-exit experiments\\
Student's t test & $t_{(52776)}$=17.882 & P$<$0.001 & 0.052 to 0.065 & 0.156 & Single-exit experiments\\
Welch's t test & $t_{(50206)}$=29.395 & P$<$0.001 & 0.057 to 0.065 & 0.143 & Single-exit experiments\\
\hline
Kolmogorov-Smirnov & $D_{(28267)}$=0.260 & P$<$0.001 & 0.089 to 0.125 & 0.138 & Exit-obscured chase\\
Student's t test & $t_{(28266)}$=11.586 & P$<$0.001 & 0.089 to 0.125 & 0.138 & Exit-obscured chase\\
Welch's t test & $t_{(25135)}$=15.966 & P$<$0.001 & 0.086 to 0.110 & 0.114 & Exit-obscured chase\\
\hline
\end{tabular}

\label{table:1}
\end{table*}

To further explore pedestrians with higher desired velocities in the presence of moving threats, the relationships between different desired velocities and their corresponding velocities and densities were examined, as shown in Fig. \ref{Fig.2}. To calculate the desired velocity curves, the Kladek-Newell-Weidmann equation (see Methods for details) was applied. The resulting curves revealed that pedestrians with higher desired velocities have higher speeds at the respective densities. An increase in density results in a decrease in the velocity of different desired velocity curves as pedestrian velocity is constrained by available space for motion. In single-exit experiments, the highest desired velocity for the moving threats condition reached 4 $\text{ms}^{-1}$, with many data points for speeds exceeding 1.5 $\text{ms}^{-1}$, while the highest desired velocity for the no-moving threats condition was found to be between 1 $\text{ms}^{-1}$ and 2 $\text{ms}^{-1}$. In general, a pedestrian walks at approximately 1.2 $\text{ms}^{-1}$ to 1.5 $\text{ms}^{-1}$; thus, it is generally suggested that the desired velocity be approximately 1.2 $\text{ms}^{-1}$ to 1.5 $\text{ms}^{-1}$ as a general baseline. This indicates that pedestrians in experiments involving moving threats achieve higher desired velocities than pedestrians in general and pedestrians in the same scenario without moving threats. In exit-obscured chase scenarios, pedestrians also achieve higher desired velocities when moving threats are present. Importantly, the speed-density data collected from experiments involving moving threats did not exhibit a higher desired velocity at a density exceeding 3 $\text{m}^{-2}$. However, pedestrian velocity data are essentially similar regardless of whether moving threats are present or absent at densities greater than 3 $\text{m}^{-2}$. This is because the space available for pedestrian movement is limited, which makes it more difficult for pedestrians to increase their speed. Consequently, valid results for the desired velocity are also difficult to obtain in this density range.

Our results indicate that pedestrians tend to achieve higher desired velocities at densities between 0 $\text{m}^{-2}$ and 3 $\text{m}^{-2}$ when faced with moving threats.

\subsection{A direct threat zone that has a significant impact on pedestrian behaviour\label{sec32}}

\begin{figure*}
  \centering
    \includegraphics[width=1\textwidth]{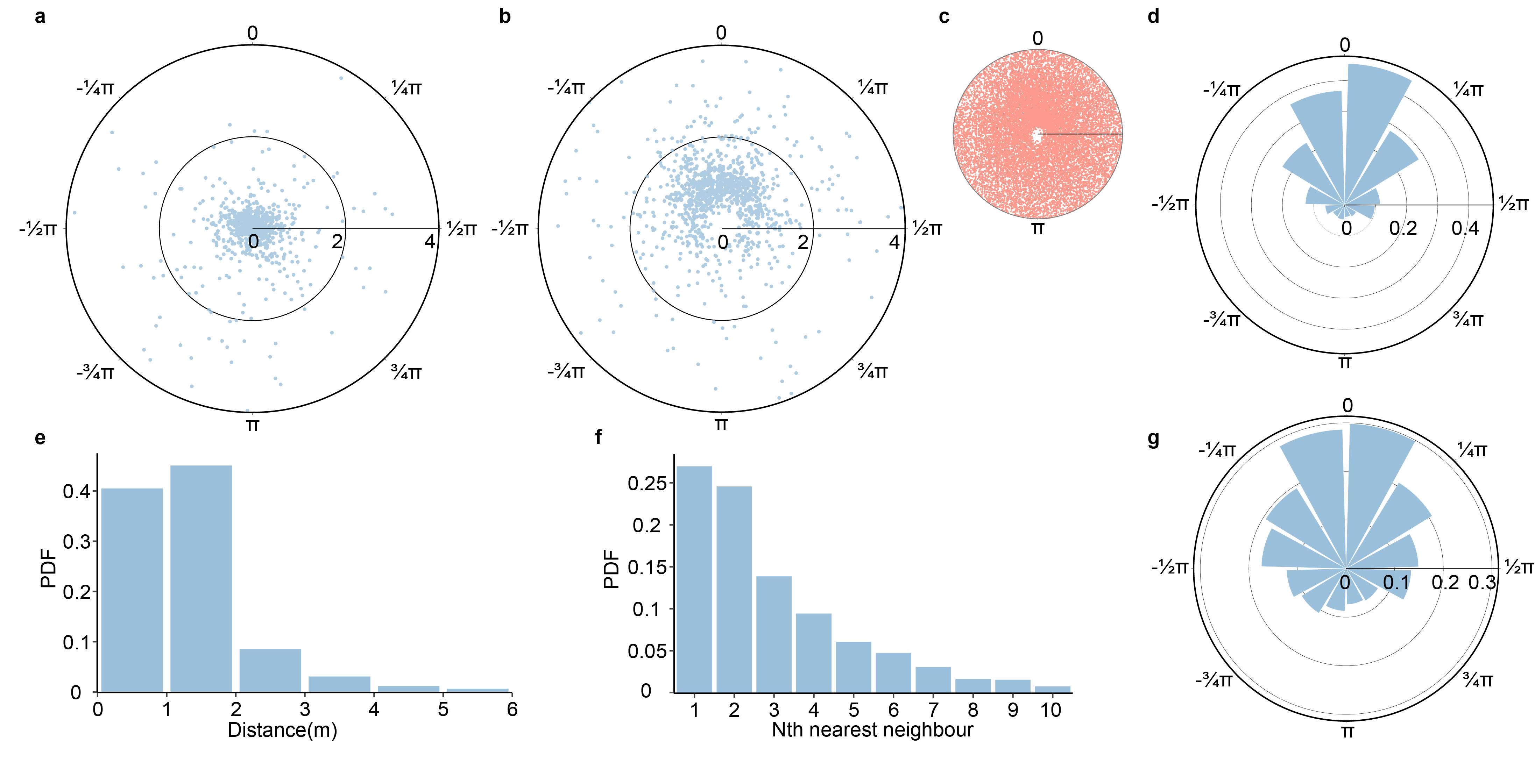}
\caption{\textbf{Assault target selection and displacement} \textbf{a}, Assault target displacement scatter plot depicting the positions where the assault targets are hit relative to their initial positions at $t_{\text{start}, i}$. The origin denotes the initial position of the target, while the dots represent the positions of the targets when hit. The angle with zero radians is identified by $\mathbf{\hat{r}}_{ap}$. The radial distance from the origin to a point reflects the distance between the hit and initial positions. \textbf{(b and c)}, The attack target selection for the attacker based on the relative orientation at $t_{\text{start},i}$. \textbf{b}, The origin indicates the attacker's position, and the scatter points represent the chosen assault targets. The radial distance is $|\mathbf{r}_{ap}|$, and the angle with zero radians is identified by $\mathbf{\hat{d}}_a$. \textbf{c} is similar to \textbf{b}, but the scatter points represent the pedestrians who are not selected. \textbf{d}, The probability density distribution of $\theta_{target}$. The blue sectors indicate regions with different target selection probabilities, with the probability scale shown on the radial axis. \textbf{e}, The histogram shows the probability density of targets being selected at varying $|\mathbf{r}_{ap}|$ at $t_{\text{start},i}$. \textbf{f}, This histogram shows the probability density of a target being selected based on its order as the nearest neighbour to the attacker. \textbf{g}, The probability density distribution of $\theta_{ap}$.
}\label{Fig.3}
\end{figure*}

To further validate our previous hypothesis, we assess how threats influence pedestrians' desired velocities. By quantitatively measuring potential threats, we aim to support our previous suggestion and elucidate how these threats impact pedestrian behaviour, focusing on the zones with the major impacts.

We investigate potential threats from the perspective of the targets selected by the attacker and define 'assault events' to analyse these selections. Given a finite sequence of assault events $S=\left\{S_i\right\}_{i=1}^N$, where each assault event $S_i$ is defined by an ordered pair of times $\left(t_{\mathrm{start}, i}, t_{\mathrm{end}, i}\right)$, the assault event sequence $S$ can be defined as:
\begin{equation}
S=\left\{\left(t_{\text {start }, i}, t_{\text {end }, i}\right) \mid t_{\text {start }, i+1}=t_{\text {end }, i} \text { for } 1 \leq i<N\right\}
\end{equation}
where $t_{\mathrm{end}, 1}$ is the last time when the weapon touches the first assault target. For each $i \geq 1, t_{\mathrm{end}, i}$ is the last time when the attacker touches the current target during the assault event $S_i$. For each $i > 1$ and $i<N$, $t_{\text {start},i }$ is equal to $t_{\text {end }, i-1}$ of the previous event $S_{i-1}$.

Each assault event involves the attacker selecting and hitting a new target after ending the previous assault. The moment $t_{start}$ marks when the attacker switches from their previous target to a new target. Our analysis of assault target displacement from $t_{\mathrm{start}, i}$ to $t_{\mathrm{end}, i}$ reveals that most selected targets have minimal movement during an assault event, as shown in Fig. \ref{Fig.3}a. This suggests that attackers may prefer to attack pedestrians with limited movement. However, the target's mobility is impeded by numerous factors, including the absence of real-time attacker location information, crowd congestion, and exit queues. Thus, identifying selection patterns from this perspective is still challenging. Therefore, we consider an alternative perspective, namely, examining the selected target from the perspective of the attacker, which reveals an apparent pattern. Concerning the distance between the target and the assailant, pedestrians located within a 2 m radius were more likely to be selected as an assault target, as shown in Fig. \ref{Fig.3}e. Furthermore, the nearest neighbour analysis revealed similar selection probabilities for the first and second nearest targets, as shown in Fig. \ref{Fig.3}f, suggesting that target selection depends on both distance and the target distribution relative to the attacker.

To examine the distribution of selected targets relative to the attacker, we must determine the direction of both the attacker's and the target's motion and their orientation at $t_{\text{start}}$. Similar to Equation (11), the attacker's and target motion vectors at $t_{\text{start},i}$ can be defined as follows:

\begin{equation}
\mathbf{v}_{a}=\frac{\mathbf{x}_{a,t-t_a+\Delta t/2}-\mathbf{x}_{a,t-t_a-\Delta t/2}}{\Delta t}
\end{equation}
\begin{equation}
\mathbf{v}_{p}=\frac{\mathbf{x}_{p,t-t_a+\Delta t/2}-\mathbf{x}_{p,t-t_a-\Delta t/2}}{\Delta t}
\end{equation}
where $\mathbf{x}_{a,t}$ and $\mathbf{x}_{p,t}$ denote the position at time $t$. $\Delta t$, represents the time interval and is set to 0.64 seconds, the same value as Equation (11). A duration of 0.64 seconds is commonly utilized to minimize deviations in the head bobbing. $t_a$ denotes the advance time needed to ensure that the direction is correct at $t_{start}$, which is set to 0.08 seconds. $t_a$ is derived according to experimental observations. Specifically, the attacker exhibits no directional alteration in trajectory within 0.24 seconds after the touch time. This ensures that the attacker's motion vectors do not prematurely align with the subsequent target. The direction of the attacker's motion is defined as $\mathbf{\hat{d}}_a = \frac{\mathbf{v}_{a}}{|\mathbf{v}_{a}|}$, where $|\mathbf{v}_{a}|$ is the magnitude of $\mathbf{v}_{a}$. Similarly, the direction of the motion of the target is defined as $\mathbf{\hat{d}}_p = \frac{\mathbf{v}_p}{|\mathbf{v}_{p}|}$. Consider $\mathbf{r}_{ap}$ as the relative position of the assault target and the attacker. The direction of the relative position is defined as $\mathbf{\hat{r}}_{ap} = \frac{\mathbf{r}_{ap}}{|\mathbf{r}_{ap}|}$, where ${|\mathbf{r}_{ap}|}$ is the magnitude of $\mathbf{r}_{ap}$. The angle between the attacker's and the target's motions is calculated as follows:
\begin{equation}
\theta_{ap} = \arccos ( \mathbf{\hat{d}}_a \cdot \mathbf{\hat{d}}_p )
\end{equation}
where $\theta_{ap}$ represents the orientation of the target motion relative to the attacker's motion at $t_{\text{start}}$ and is utilized in Fig. \ref{Fig.3}g. Fig. \ref{Fig.3}g suggests that attackers are more likely to select pedestrians who are moving in the same direction and have their backs to the attackers. Such pedestrians not only movve in alignment with the assailant's motion direction but also lack immediate awareness of the assailant's position. Although this experiment was not specifically designed to determine whether this was a direct or an indirect factor coupled with other crowding factors, it suggests a possible target selection pattern based on the distribution of targets relative to the attacker's motion direction. 
The angle between the attacker's motion direction and the relative position direction is calculated as follows:
\begin{equation}
\theta_{target} = \arccos ( \mathbf{\hat{d}}_a \cdot \mathbf{\hat{r}}_{ap} )
\end{equation}\label{eq_theta}
where $\theta_{target}$ is the orientation of the target relative to the attacker's motion at $t_{\text{start}}$.
Zones with higher concentrations of potential targets are considered to pose greater threats to pedestrians. The target distribution was analysed to identify the attacker's threat zone, as depicted in Fig. \ref{Fig.3}b and d. The distribution of the assault target relative to the attacker position is concentrated in the region where $\theta \in [-\frac{3\pi}{4}, \frac{3\pi}{4}]$. Less targets are distributed in the regions where $\theta \in [-\pi, -\frac{3\pi}{4}]$ and $\theta \in [\frac{3\pi}{4}, \pi]$. The target selection pattern depends on the pedestrian's distribution relative to the attacker's motion direction. Similarly, the potential threat to pedestrians varies according to their respective distributions.

As observed in the concentration trend of assault target distributions, there seems to be an area with a high potential threat. Considering the difficulty of the attacker's movement to a certain location, it is easier for an attacker to initiate an assault by continuing his or her direction of motion; however, it is more difficult for an attacker to initiate an assault in the opposite direction. An attacker's ease of initiating assaults depends on the needed angle change from their motion direction. Based on this assumption, we can define the boundary of direct threat zone $B(\theta)$ as follows:

\begin{equation}
  B(\theta) =
  \begin{cases}
    {\bar v_0}{\tau}\exp\left(-\frac{\theta^{2}}{2\alpha^{2}}\right)+\frac{b}{\sqrt{1-\left(\epsilon^{\alpha}\cos\left(\theta\right)\right)^{2}}},  &\theta \in [-\frac{3\pi}{4},\frac{3\pi}{4}] \\
    \\
    L_{b},   \theta  \in [-\pi,-\frac{3\pi}{4}) \cup (\frac{3\pi}{4},\pi)
  \end{cases} \label{eq1}
\end{equation}
Here, $\bar v_0$ denotes the average desired speed considered in the direct threat zone, where $\bar v_0 = 1.2$ m/s for the attacker. It is generally assumed that 1.2 m/s is the average desired speed in a pedestrian system. $\tau$ denotes the characteristic time, indicating that only impacts caused by direct threats during the $\tau$ time period are considered. Here, we set $\tau$ = 1 s. $\alpha$ is the moving threat manoeuvring rate, which measures the ability to manoeuvre from the direction of motion to the side and rear during the characteristic time. As the manoeuvrability $\alpha$ increases, the boundary of the direct threat zone becomes wider. There is a longer direct threat range on the sides. As $\alpha$ decreases, the direct threat range becomes more concentrated in the direction of motion. $\alpha$ = 0.85 in the results fitted to the distribution of the assault target. $\epsilon$ represents the eccentricity of the moving threat shape. The aspect ratio of the shape for moving threats, such as vehicles or robots, may not be equivalent to one. Importantly, in this case, the attacker is represented by a circle with $\epsilon$ = 0. The semiminor axis of the static threat ellipse is denoted by $b$. For the attacker, we set $b$ = 0.85 m. This distance includes the 0.6-m length of the weapon stick, along with the average 0.25-m length of a participant's upper arm. $L_{b}$ is the length of the back half of a moving threat. In this case, the attacker circle has a radius of 0.2 m. Here, the value of $L_{b}$ is 0.2 m; thus, $L_{b}$ = 0.2 m. In the frontal 270-degree field, a Gaussian-like function is used to calculate $B(\theta)$; otherwise, the value of $B(\theta)$ is the length of the back half of a moving threat.

Equation (\ref{eq1}) describes the boundary of the direct threat zone. The direct threat zone is defined in terms of $r^{dr}$ and $\theta$ with respect to the boundary $B(\theta)$. The direct threat zone is described as follows:
\begin{equation}
z_{dr} = \{ (r^{dr}, \theta) \mid 0 \le r^{dr} < B(\theta),\ -\pi \le \theta < \pi \} \label{eq2}
\end{equation}
Here, $r^{dr}$ is the radius of the direct threat zone, and $\theta$ is the angle of the pedestrian relative to the direction of motion of the attacker. When $\theta$ is 0, the pedestrian is moving in the same direction as the attacker.

Equation (\ref{eq2}) describes a direct threat zone, i.e., the zone where nonreciprocal effects have the greatest impact on pedestrians. The equation can describe the impact of mobile threat nonreciprocal effects on pedestrian behaviour at the system level. It can likewise be described at the micro level. When considering the micro level, $\bar v_0$ is replaced by the instantaneous speed. If the instantaneous speed is 0, then the region of maximum impact of the moving threat on the pedestrian is in its static threat range (the attacker is considered a circle with a radius of 0.85 m from the distance of the arm holding the stick). When the instantaneous speed is high, pedestrians at longer distances from the moving threat movement distance are affected. This kind of description also applies to the nonreciprocal effects of moving threats such as attackers, vehicles, and robots.

\begin{figure*}[ht]
  \centering
    \includegraphics[width=1\textwidth]{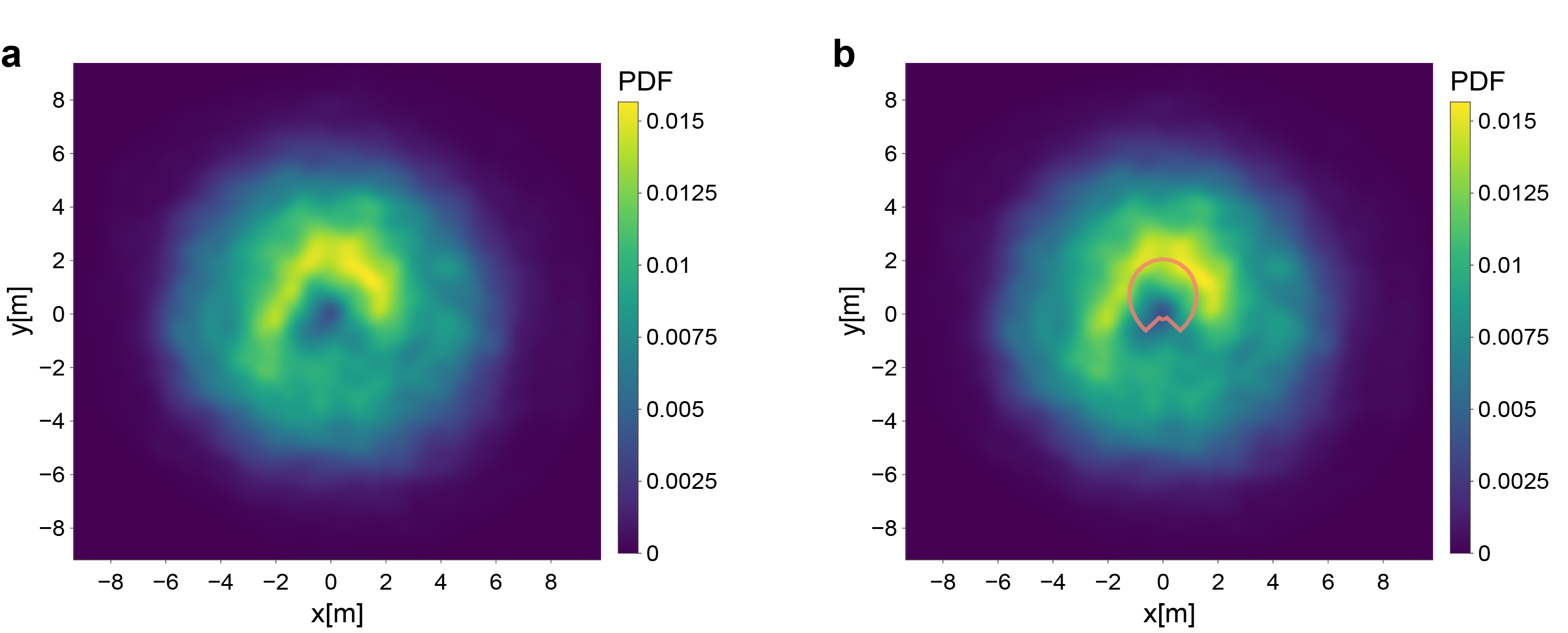}
\caption{\textbf{Probability density function for the distribution of the pedestrian's motion position relative to the attacker.} These figures represent the relative positions of pedestrians in relation to the attacker at each moment in the three experiments. Additionally, the statistics do not include any casualties because they will remain in place as required \textbf{a}. A semicircular pattern emerges in which fewer pedestrians are in the attacker's attack-blind field but are concentrated at a farther distance. The semicircular pattern is primarily a result of the different interactions between the attacker and the pedestrian. The attacker tends to cover the pedestrian travelling path within the direct threat zone and thereby restricts the pedestrian motion space. Pedestrians who are affected by a chasing attacker tend to move along the edge of the direct threat zone to avoid falling into that zone. As a result, pedestrians appear at the edge of the direct threat zone more frequently. \textbf{b}, A comparison of the semicircular pattern and the attacker's direct threat zone. The orange line indicates the boundary of the attacker's direct threat zone $r^d = B(\theta)$, and the inside indicates the direct threat zone $r^d < B(\theta)$. The semicircular pattern corresponds to our definition of the direct threat zone. This confirms the validity and objectivity of the direct threat zone proposed in this study.}\label{Fig.4}
\end{figure*}

To further validate the inferred direct threat zone, we counted assault targets within the direct threat zone. Over 70\% of the assault targets were located within the direct threat zone according to various experiments. The direct threat zone covered the majority of the assault target distribution, as shown in Fig. \ref{Fig.6}b. This finding shows that zone division is meaningful for describing potential threats. Due to the importance of direct threat zones, it should be ensured that the analyses are free of empirical or statistical bias. To verify this, we analysed the issue from another perspective. Due to the presence of direct threat zones, it is natural to speculate that there was less pedestrian movement inside the area. Furthermore, pedestrians will more often be found at the edge of the direct threat zone as the attacker tries to cover the pedestrian evacuation path or pedestrian movement space within the direct threat zone. This causes pedestrians to move sideways along the direct threat zone to bypass it. Fig. \ref{Fig.4} shows the distribution of the relative positions of the two teams during the confrontation. The overall distribution is close to a circle, with a half-ring pattern in the zone close to the moving threat. The pattern is formed mainly by the pedestrian's avoidance and the attacker's chasing. When an attacker chases a pedestrian, the attacker chooses the pedestrian who is likely to be hit, usually one who is close to the direct threat zone or restricted in movement. This implies that the target has insufficient time and space for movement and will instead move directly away from the attacker. Thus, the best strategy for a pedestrian in the neighbourhood of a chase target is to move along the tangent of the attacker's direct threat zone when the chase target is moving away to reach a safer blind spot behind the attacker. This behaviour can also be observed in a real case (Appendix, Fig. \ref{figa1}). These findings may confirm that a direct threat zone exists and are consistent with the empirical parameters that we set for this zone.

We propose an equation to represent the direct threat zone range. The scale of the direct threat zone is determined by the estimated movement speed of the moving threat along with the direct range of damage. Furthermore, the relative position distribution during the confrontation confirms the existence of a direct threat zone.

\subsection{The impact of potential threats may cause a different desired velocity\label{sec33}}

\begin{figure*}
  \centering
      \includegraphics[width=1\textwidth]{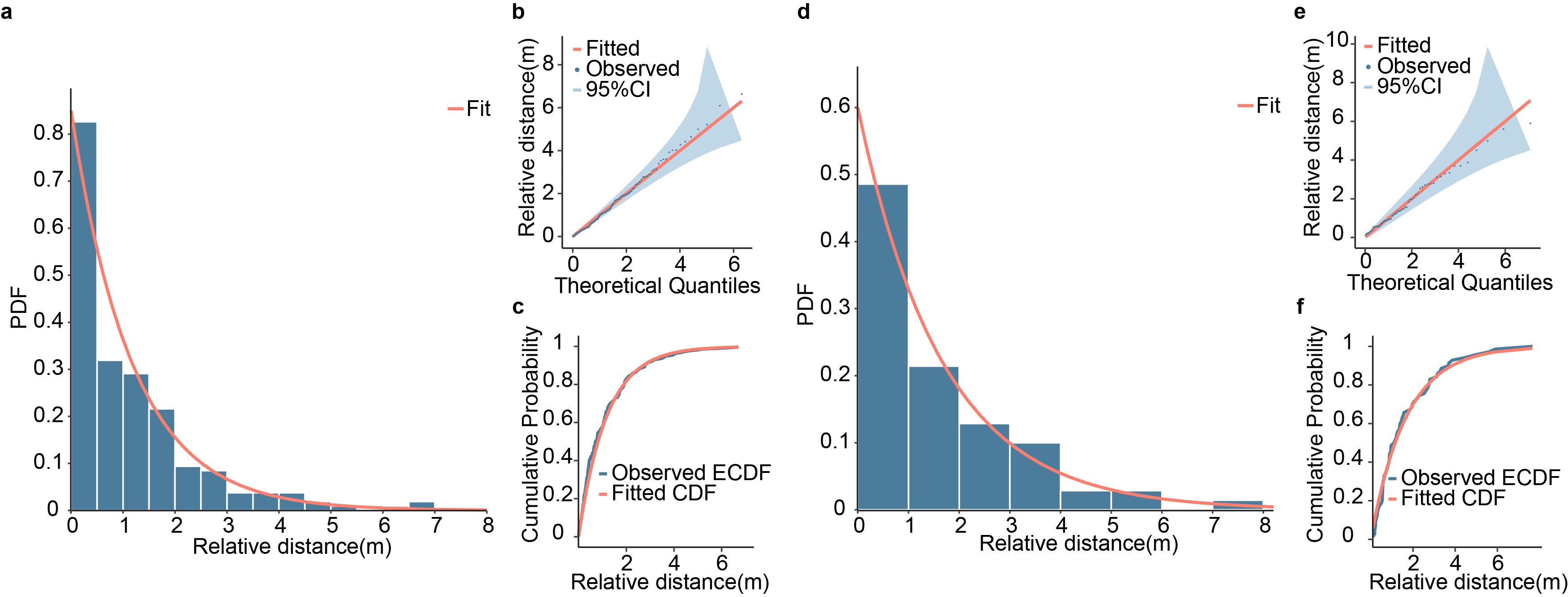}
\caption{\textbf{Quantification of potential threats and validation of quantified distributions.} \textbf{a}, Histogram of the distance distributions for assault targets in adjacent threat zones relative to direct threat zones. The fitted distribution function is exponential, $f(r_{ad}) = 0.85exp(-0.85 r_{ad})$. $r_{ad}$ is the relative distance between the assault target and the direct threat zone boundary. \textbf{b}, Quantile-quantile plot of the fitted distribution of the adjacent threat zone. The observed data are basically distributed over the fitted distribution. \textbf{c}, The empirical cumulative distribution function of the fitted distribution of the adjacent threat zone compared to the observed data distribution. The fitted distribution matches the empirical cumulative distribution function of the observed distribution. As a result of \textbf{(b-c)}, it can be concluded that $f(r_{ad})$ can accurately reflect the actual distribution of data in the adjacent threat zone. \textbf{d}, Histogram of the distance distributions of the assault targets in the rearview threat zone. The fitting function is an exponential distribution with $g(r_{rv}) = 0.6\exp(-0.6 r_{rv})$. $r_{rv}$ is the relative distance between the assault target and the body of the attacker (a circle with a radius of 0.2 m, and the centre of the circle is the attacker's location). \textbf{e}, Quantile-quantile plot of the fitted distribution for the rearview threat zone. The observed data distribution conforms to the fitted distribution. \textbf{f}, The empirical cumulative distribution function of the fitted distribution for the rearview threat zone compared with the real observed distribution. The fitted distribution differs only slightly from the distribution of the observed data. According to \textbf{(e-f)}, We can conclude that $g(r_{ad})$ is a reliable estimate of the observed data distribution in the rearview threat zone.}\label{Fig.5}
\end{figure*}

\begin{figure*}
  \centering
      \includegraphics[width=1\textwidth]{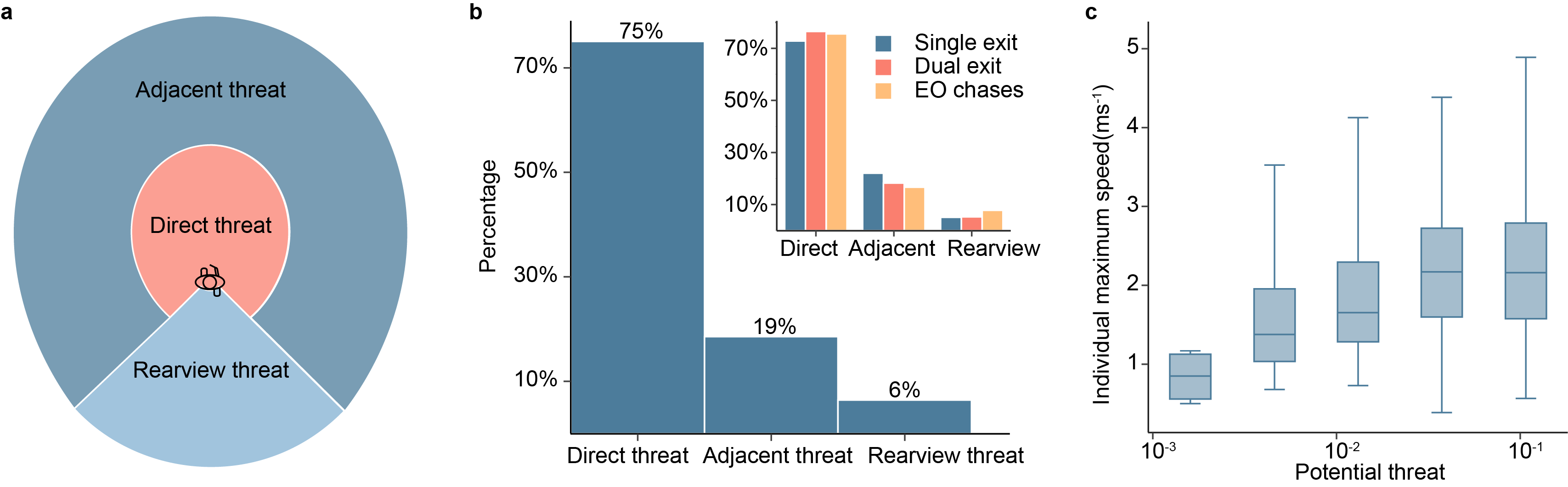}
\caption{\textbf{Impact of different potential threats on desired velocity.} \textbf{a}, Schematic diagram of the potential threat zones divided into three zones: a direct threat zone, an adjacent threat zone, and a rearview threat zone. A schematic diagram of the division of these three threat zones is similar to an ellipse. \textbf{b}, The percentage of assault targets in different zones. The percentage is calculated by dividing the number of attack targets within each zone by the total count of attack targets in each respective scenario. In all the experimental scenarios, the proportion of assault targets in the direct threat zone exceeded 70\%, for an overall percentage of 75\%. The assault target percentages of the different potential threat zones in the different experiments were similar, except for the rearview zone percentage in the exit-obscure chase experiment. \textbf{c}, Personal maximum speed vs. potential threat. Individual maximum speed and potential threat were obtained at the moment of individual maximum speed in each experiment. The potential threat can be calculated by Equation (\ref{eq3}) considering the relative distance and angle of a pedestrian at that moment.}\label{Fig.6}
\end{figure*}

In addition to direct threat zones, assault targets are also distributed in other zones. Thus, we divided the assault target distribution zones into a direct threat zone, an adjacent threat zone, and a rearview threat zone, as shown in Fig. \ref{Fig.6}a. The adjacent threat zone refers to the area that is outside the direct threat zone but not part of the blind area of the field of view. The adjacent threat zone is defined as follows:
\begin{equation}
z_{ad} = \{(r^{ad}, \theta) \mid r^{ad} \ge B(\theta),\ \theta \in \left[0,\frac{3\pi}{4}\right) \cup \left(\frac{5\pi}{4},2\pi\right) \} \label{eq_ad}
\end{equation}

The rearview threat zone is the zone located on the back side of the motion direction. The rearview threat zone is defined as follows:

\begin{equation}
z_{rv} = \{(r^{rv}, \theta) \mid r^{rv} \ge B(\theta),\ \theta \in \left[\frac{3\pi}{4},\frac{5\pi}{4}\right]\} \label{eq4}
\end{equation}

We fitted the probability densities of the assault target distributions in the adjacent threat zone and rearview threat zone. The fitted probability distributions were exponential, as shown in Fig. \ref{Fig.5}. Furthermore, hypothesis testing was performed for the distribution in each zone with the corresponding exponential distribution.

The Kolmogorov-Smirnov test and Student's t test confirmed that the distributions within the adjacent threat zone and rearview threat zone were not significantly different from the corresponding exponential distributions. For the adjacent threat zone, the Kolmogorov-Smirnov test (D(425)=0.085; two-tailed P=0.433; 95\% confidence interval (CI)=-0.224 to 0.249; Cohen's d=0.010) and Student's t test (t(424)=0.106; two-tailed P=0.916; 95\% CI=-0.224 to 0.249; Cohen's d=0.010) confirmed that the distribution within the adjacent threat zone was not significantly different from the corresponding exponential distribution $f(r_{ad})$. For the rearview threat zone, the Kolmogorov-Smirnov test (D(139)=0.129; two-tailed P=0.613; 95\% confidence interval (CI)=-0.464 to 0.594; Cohen's d=0.041) and Student's t test (t(138)=0.244; two-tailed P=0.808; 95\% CI=-0.464 to 0.594; Cohen's d=0.041) confirmed that the distribution within the rearview threat zone was not significantly different from the corresponding exponential distribution $g(r_{rv})$. According to the distribution, the lambda for the adjacent threat zone was 0.85, and the expected value was 1.18. The lambda for the rearview threat zone was 0.6, and the expected value was 1.67. These results indicate that pedestrians in the blind zone of view are less likely to be threatened at the same relative distance from the direct threat zone.

The equation defines a function $p(r_{ij}, \theta)$ that describes three distinct regions characterized by the angle $\theta$ and the distance $r_{ij}$. Based on the assault target probability percentages of the three zones in Fig. \ref{Fig.6}b, the following potential threats can be calculated:

\begin{equation}
  p(r_{ij}, \theta) = 
  \begin{cases}
    h_d,&(r_{ij},\theta) \in z_{dr}\\
    h_{ad} \lambda_{ad} e^{-\lambda_{ad} (r_{ij} - B)}, & (r_{ij}, \theta)  \in z_{ad} \\
    h_{rv} \lambda_{rv} e^{-\lambda_{rv} (r_{ij} - r_b)}, & (r_{ij}, \theta)  \in z_{rv}
  \end{cases} \label{eq3}
\end{equation}

where this equation describes the potential threat degree in different zones. $h_{dr}$, $h_{ad}$ and $h_{rv}$ represent the percentages of attack targets in different zones, and their values are 0.75, 0.19 and 0.06, respectively. These values are used as weights for different zones to adjust the potential threat degree in different zones. $\lambda_{ad}$ and $\lambda_{rv}$ denote the fitted parameters for the distribution of attack targets with respect to their zones, as shown in Fig. \ref{Fig.5}, with $\lambda_{ad}$ = 0.65 and $\lambda_{rv}$ = 0.8. $r_{ij}$ is the relative distance of pedestrians from the attacker. $r_b$ is 0.2 m, which is the radius of the attacker's body.

In our previous section, we noted the need for a quantified degree of potential threat to further illustrate desired velocity variations. Thus, we categorized the potential threats into different threat zones and quantified the potential threats to pedestrians. A pedestrian's desired velocity is typically defined as the maximum speed at which he or she expects to travel when unconstrained. As the actual situation may not always be unconstrained, a collected individual's maximum speed represents a lower bound on his or her desired velocity. The desired velocity is typically greater than or equal to an individual's maximum speed recorded in the data. Fig. \ref{Fig.6}c shows a representation of the pedestrian's maximum speed in relation to potential threats. The pedestrian's maximum speed was found to increase with the potential threat until it reached the maximum speed and then remained stable. Pearson's correlation analysis (correlation coefficient: 0.283, P $<$ 0.001) and Spearman's correlation analysis (correlation coefficient: 0.291, P $<$ 0.001) confirmed a statistically positive relationship between the maximum speed of the pedestrian and the potential threat to the pedestrian. In other words, the desired pedestrian velocity was significantly positively correlated with the severity of the potential threat. The individual's maximum speed increased with the potential threat until the potential threat reached approximately 0.1, at which point the increase in desired velocity tended to slow. This also means that the desired velocity of the pedestrian continued to increase until it reached 0.57 m from the direct threat zone, except in the attacker's blind zone. This also explains the change in the desired velocity of pedestrians found in the study of the Running of the Bulls Festival in Pamplona, Spain \citep{parisipedestrian2021}; i.e., pedestrians exhibited a high desired velocity when they were near the direct threat zone of moving threats.

This result, along with our previous speed-density fundamental diagram results, demonstrates that the potential threat affects pedestrian motion behaviour, particularly by increasing their desired velocity. Moreover, the desired velocity increases when the pedestrian is closer to the direct threat zone of the moving threat.

\subsection{Characteristics of the steady state in a crowd confrontation system\label{sec34}}

\begin{figure*}
  \centering
      \includegraphics[width=1\textwidth]{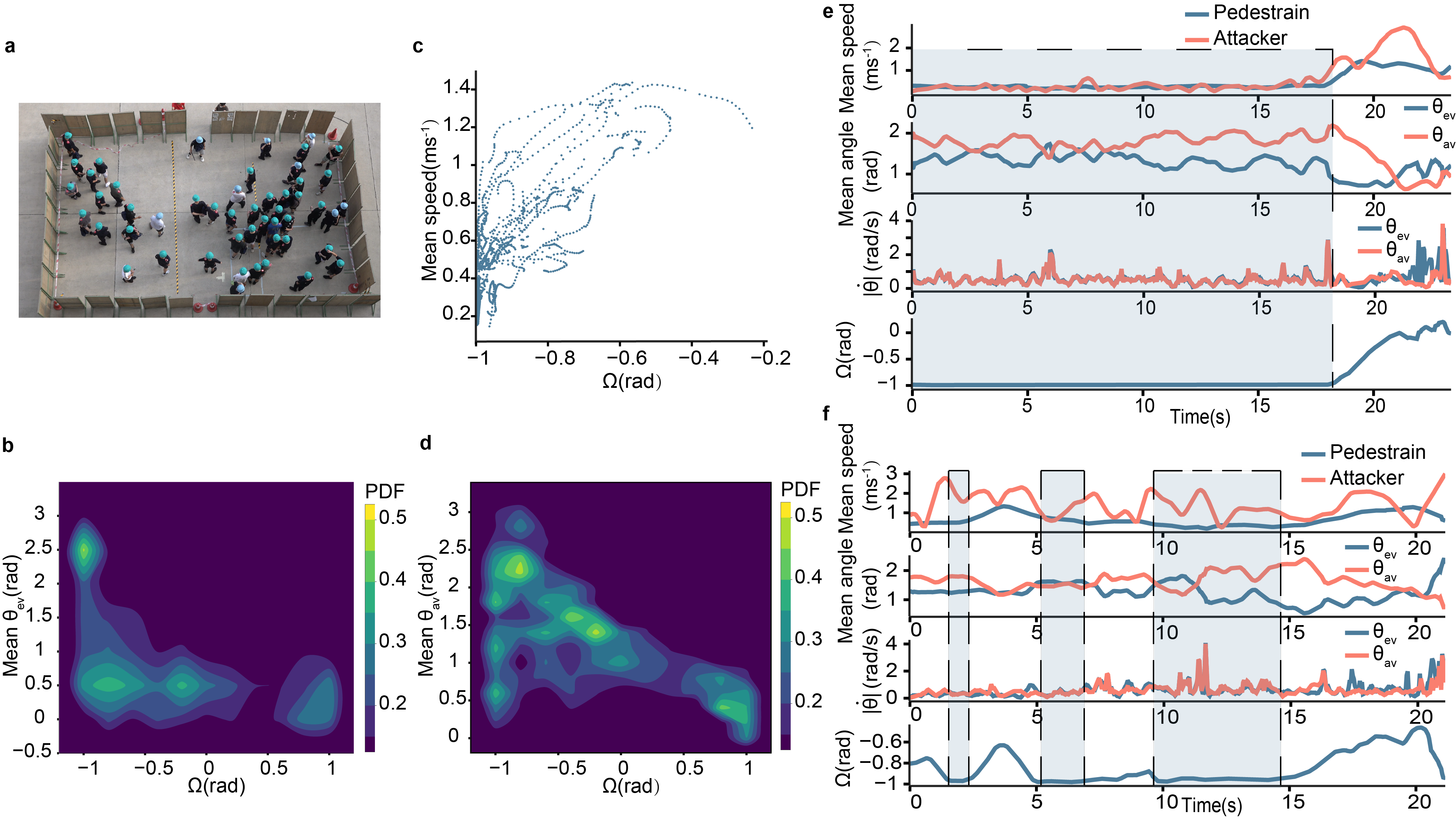}
\caption{\textbf{The characteristics of a steady state in nonreciprocal interaction crowd systems.} \textbf{a}, A screenshot of the steady state as observed in the experiment. The pedestrian and the attacker are in a standoff in this state, and the pedestrian leaves a safe space to observe further movement before engaging in their own movement. \textbf{(b,d)}, Density heatmaps of $\Omega$ versus the angle between the desired direction and motion direction. The mean desired directions are $\theta_{ev}$ (approaching the exit) and $\theta_{av}$ (avoiding the attacker), where $\omega = \cos(\theta_{ae})$, $\Omega$ represents the mean value of the crowd $\omega$, and $\theta_{ae}$ is the angle between the two desired directions. $\Omega$ represents the degree to which the attacker blocks the exit. $\Omega = -1$ represents a complete blockade. $\Omega = 1$ indicates that the attacker is located behind the exit crowd. The heatmaps show that when the exit is completely blocked, the direction of the crowd motion is not direct. As the exit blockade begins to loosen, pedestrians will preferentially head for the exit rather than avoid the attacker. \textbf{(e-f)}, Time series of the mean speed for different teams, the mean angle between different desired directions and the pedestrian's motion direction, and the degree to which the exit is blocked by the attacker in two typical single-exit experiments. The time series are smoothed by the Wiener filter with an eight-frame length window \citep{norbertextrapolation1949}. \textbf{e}, The attacker completely blocks the exit from the beginning of the experiment, and the speed of both teams approaches zero, indicating a vanishing pedestrian's desired evasion velocity. \textbf{f}, The attacker switches between actively attacking and blocking the exit. Similarly, when the attacker completely blocks the exit, the pedestrian's desired evasion velocity vanishes. In \textbf{(e-f)}, As the attacker completely blocks the exit, the absolute value of the angular change rate of crowd motion with respect to the different targets is almost identical.}\label{Fig.7}
\end{figure*}

While traditional pedestrian models refer to a "desired speed" under obstacle-free conditions, in the context of the present study, in which an external threat is present, we introduce the term "desired evasion velocity". The desired evasion velocity represents the velocity a pedestrian would aim for in the presence of a threat, accounting for the desire to approach the safety area and escape from the threat. Our observations from the experiments also suggest that the desired evasion velocity of pedestrians is more complex than initially understood. There was a steady state in the experiment where the pedestrian's desired evasion velocity disappeared, as shown in Fig. \ref{Fig.7}a. The pedestrian velocity in this situation was found to be almost stagnant, despite the low density of people around the pedestrian. Thus, the speed-density relationship for pedestrians deviates from the general relationship for pedestrian dynamics. In addition, this steady state was also observed in a real case (Appendix, Fig. \ref{figa2}). The necessary condition for the emergence of a steady state seems to be that the direct threat zone of the moving threat blocks the effective path for pedestrian evacuation. Thus, we next examine the characteristics of the pedestrian system in this steady state from the perspective of the blockade of effective evacuation paths for pedestrian evacuation.

It is assumed that pedestrian motion in the system has two principal subgoals, namely, approaching the exit and avoiding threats. A pedestrian expects to maximize the total goal advantage by accumulating a microscopic advantage at each movement step. The overall goal comprises the microscopic advantages gained by approaching the exit and avoiding the threat. We define the unit vectors of the direction for different pedestrian motion goals, namely, \textit{e} (pedestrian evacuation direction) and \textit{a} (pedestrian avoidance direction), to describe the desired motion direction of the pedestrian for different goals. $\bf{e} = \frac{x_e - x_p}{\lVert x_e - x_p \rVert}$, $\bf{a} = \frac{x_p - x_a}{\lVert x_p - x_a \rVert}$, where $\bf{x_e}$ is the exit midpoint, $\bf{x_p}$ is the pedestrian position, and $\bf{x_a}$ is the attacker position. We define a simplified measure $\omega$ to describe the blockage of effective pedestrian evacuation paths generated by the threat. $\omega = \cos(\theta_{ae})$, where $\theta_{ae}$ is the angle between the two desired directions.

To further study crowd behaviour, we define $\Omega$ as the mean $\omega$ of the crowd, where $\Omega\in [-1,1]$. $\Omega=-1$ indicates that the moving threats completely block the crowd's direction to the exit. For the pedestrian, the two motion targets are in opposite directions. A greater $\Omega$ indicates a more relaxed blockade. When $\Omega$ is close to -1, the mean crowd speed is slow and is the local minimum in the time series. The duration of the steady state depends on the period for which $\Omega$ is close to -1.
In this state, there is a broad motion space around the pedestrian, but the speeds are low. This is considered evidence of the disappearance of the desired evasion velocity. Additionally, as $\Omega$ approaches -1, the crowd appears to be synchronized in the direction of motion for both targets. The angular change rate of crowd motion with respect to the different targets has an almost identical absolute value, as shown in Fig. \ref{Fig.7}(e-f). This is one of the crowd behaviours that emerges when the exit direction is completely blocked by moving threats. Given that the two desired direction vectors are nearly opposite at this point, moving in either direction does not lead to a significant increase in the total target advantage. Furthermore, in Fig. \ref{Fig.7}c, it can be observed that the crowd speed is quite slow in situations of full blockade. As $\Omega$ moves away from -1, the crowd speed begins to recover and grow rapidly until $\Omega = -0.6$. Fig. \ref{Fig.7}b and Fig. \ref{Fig.7}d show the angles between the direction of crowd motion and the two desired directions at different blockade levels. There is no specific directionality between the two desired directions when moving threats completely block the exit path. However, pedestrian motion exhibits directionality when the blockade is gradually relaxed. Pedestrians are more likely to move towards the exit direction than directly away from the threat. As the blockade of the moving threats becomes more relaxed, pedestrians can increase the advantages for both subgoals at once through a direction of motion. Then, the pedestrian motion direction has a strong level of determinism.

Our observations suggest that a steady state may emerge when the approach direction of a pedestrian's exit is completely blocked by moving threats, resulting in the disappearance of the desired evasion velocity.

\section{Discussion\label{sec4}}

In this study, we investigate nonreciprocal crowd dynamics via experiments simulating mass stabbing. Specifically, the effect of strongly coupled nonreciprocal interactions on the pedestrian's behaviour, and on the pedestrian's desired velocity. Herein, we summarize our major findings and discuss their implications for pedestrian dynamics and future research directions.

Our experiments included single-exit, dual-exit, and exit-obscured chase scenarios. We observed that pedestrians tend to achieve higher desired velocities when confronted with moving threats at densities ranging from 0 $\text{m}^{-2}$ to 3 $\text{m}^{-2}$. Furthermore, we defined an equation for the direct threat zone that most affects pedestrian behaviour. This equation revealed the factors that determine the extent of the direct threat zone. Then, we identified two distinct threat zones, namely, an adjacent threat zone and a rearview threat zone. These findings led us to propose a potential threat equation, which showed a positive correlation between the desired pedestrian velocity and the degree of potential threat. Similarly, we can observe the effect of the magnitude of strongly coupled nonreciprocal interactions on pedestrian behaviour from these.

The speed-density fundamental diagrams revealed that the presence of moving threats led to a higher desired velocity and a distinct speed-density relationship in our experiments. This observation is consistent with the real-life violence videos reported by \cite{bernardiniterrorist2021} and the Running of the Bulls Festival in Pamplona, Spain \cite{parisipedestrian2021}, suggesting that the findings are applicable to real-world scenarios. In addition, our proposed direct threat zone was supported by the distribution of assault targets and the distribution of pedestrians' relative motion positions, further validating its significance. Finally, our potential threat equation indicated that the desired velocity of a pedestrian increased as the potential threat increased until a certain position was reached near the edge of the direct threat zone. This outcome revealed the transition mechanism of the desired velocity. In addition, it further verified that the presence of moving threats is the reason for the higher desired velocity and the distinct speed-density relationship. Furthermore, we observed a special case in which the desired velocity disappeared (a steady state in the nonreciprocal interaction crowd system). We described two desired directions of pedestrian motion to provide an explanation for the emergence of this steady state and the characteristics of pedestrians for this duration.

In the previously mentioned study of the Running of the Bulls Festival in Pamplona, Spain, \cite{parisipedestrian2021} assumed that in scenarios with moving threats, pedestrians consistently exhibit a higher desired velocity, regardless of density. Our findings partially support this assumption. However, we cannot definitively confirm this hypothesis at all densities. For safety reasons, the fundamental diagrams cannot reach the "falling" areas described in the study. We suspended the experiment if a pedestrian fell during the experiment. Thus, although we have a limited number of samples supporting the validity of the argument at densities less than 3 $\text{m}^{-2}$, the evidence remains inconclusive. Our findings build upon and extend the current understanding of pedestrian dynamics. The identification of threat zones and the proposed potential threat equations provide a more detailed understanding of pedestrian decision-making and movement behaviour under the impact of strongly coupled nonreciprocal interactions with moving threats.

Despite our new contributions, this study has several limitations. First, our experiments were conducted in a controlled environment, which may not entirely replicate the complexity of real-world scenarios. For example, when the density in front of the exit is high, participants normally push each other; however, for safety reasons, we did not suggest that the participants push each other to compete for the exit in the experiment. This may have led to pedestrian behaviour in our study differing from that in reality at a density greater than 3 $\text{m}^{-2}$. Second, a sex imbalance arose partly due to the availability of volunteers. Specifically, there were fewer female volunteers due to the topic of the experiment. Individuals were not forced to participate in the experiment. Therefore, we tried our best to ensure that both genders were represented by our participants and to recruit as many female volunteers as possible. The final male-to-female ratio was not an intentional choice, and this factor limits the generalizability of the study to broader populations. Gender differences could play a significant role in crowd dynamics and behaviours, especially during emergencies. Future research should include verifying these findings in more diverse contexts and larger sample sizes. Third, the potential threat equation presented in this study is a simplified representation of pedestrian behaviour and should be further refined through additional data, including the effects of individual decision-making or other environmental factors. Fourth, we assume that the attacker's intention remains unchanged throughout the entire attack event. This assumption might lead to biases in the target selection analysis. Although we corrected some of the target choices manually, there may still be a few cases in which the data were unobservable. In future research, additional observational methodologies will be implemented to conduct specialized experiments. Fifth, the "threat" generated by the attacker and the "escape from the attacker" by the pedestrians mentioned in our study are strongly coupled, which is typical of nonreciprocal interactions. However, strongly coupled nonreciprocal interactions are not the only nonreciprocal interactions. Thus, the results obtained in our study support the effect of only strongly coupled nonreciprocal interactions. Our results partially improve our understanding of nonreciprocal interaction crowd systems. We plan to investigate additional kinds of nonreciprocal interactions in populations in the future to further improve our understanding of nonreciprocal interaction crowd system dynamics.

\section{Conclusion\label{sec5}}

Our findings provide new insights into nonreciprocal interaction crowd dynamics and the impact of moving threats on pedestrian dynamics. The speed-density fundamental diagram with quantitative potential threat analysis points to a higher desired speed for pedestrians in strongly coupled nonreciprocal interactions with moving threats. In future studies of crowd dynamics involving strongly coupled nonreciprocal interactions, the effect of this desired speed instead of referring to previous pedestrian dynamic speed-density patterns can be considered. The proposed equation for a direct threat zone explains that the major effect of a moving threat on pedestrian behaviour depends on the motion velocity of the moving threat and the radius of the hazard generation zone. Furthermore, future research can explore direct threat zones associated with other moving threats, such as vehicles and robots. Different moving threats may have different parameters that affect the extent of their direct threat zones. The steady state of nonreciprocal interaction crowd dynamics represents a self-organizing phenomenon of crowd wisdom under confrontation; this phenomenon can be further studied in the field of crowd intelligence.

In a practical sense, our results can provide guidance to managers and individuals regarding reducing casualties in mass stabbings. Our results illustrate the existence of a direct threat zone for a moving threat. The pedestrian's desired velocity is greater when there is a moving threat, and the desired velocity is lower when the pedestrian is far from the direct threat zone. It is suggested that pedestrians always maintain a high desired velocity instead of reaching their maximum desired velocity when approaching a direct threat zone. This approach involves helping other pedestrians move away from the threat as quickly as possible while allowing others to have more space in which to move and reducing the density of the room. This can allow the other pedestrians to reach speeds as close to their desired velocity as possible even at medium to high densities. Due to attacker-pedestrian interactions, pedestrian movement tends to concentrate more at the boundary of the direct threat zone. Pedestrians should move directly away from the direct threat zone to avoid falling into it. If they are inevitably located within the direct threat zone, they should move along the boundary of the direct threat zone towards the attacker's rearview zone and look for opportunities to reach a safe area. As the crowd-attacker system reaches the steady state (standoff), pedestrians should wait for the attacker to act before deciding, similar to crowd wisdom that emerges in this state. This is because an attacker is always constrained by the timing of an attack (note that this study did not consider hostage-taking scenarios). The appearance of the steady state may be due to the attacker blocking the pedestrian evacuation route. Therefore, we suggest that indoor spaces be designed with multiple exits (we did not observe this steady state in a single-attacker dual-exit experiment). Placing exits in opposite directions, rather than in the same direction, can also help avoid conflicts in pedestrians' desired directions. Such conflicts can lead to a reduction in or even the disappearance of pedestrians' desired velocity. In future laboratory experiments, it should be investigated whether the difference in casualty rates between scenarios with multiple exits in the same direction and opposite directions.

In conclusion, our findings provide new insights into the impact of moving threats on pedestrian dynamics and may offer a humble starting point for future experimental studies of nonreciprocal interaction crowd dynamics. These findings may have implications for designing evacuation plans, such as the positioning of exits in various buildings and urban spaces. In addition, these findings may have implications for improving individual self-protection during mass stabbing.

\section{Methods\label{sec6}}
\subsection{Statistical methods}
In Subsection \ref{sec31}, we employed a variety of statistical tools. These tests included the Kolmogorov-Smirnov test, Student's t test, and Welch's t test. These tests were used to establish a statistically significant distribution. Specifically, we examined the velocity dispersion for single-exit and obscured-exit chases experiments. The use of these tests facilitated a multifaceted approach towards hypothesis evaluation. The Kolmogorov-Smirnov test is a nonparametric measure. In the model, we refrain from making assumptions regarding the data distribution. This contrasts with the Student's t test, which is parametric. The data were assumed to be normally distributed. Welch's t test is a modified version of Student's t test. Assumptions are not made about equal variances. By using these three tests, we enhanced the robustness of our findings. Our hypothesis was confirmed under diverse conditions and assumptions.

In Subsection \ref{sec33}, we utilized the Kolmogorov-Smirnov test and Student's t test. These tests were used to ascertain that the distributions within specific threat zones did not statistically deviate from the corresponding exponential distribution. The use of both tests improved the robustness of our hypothesis testing. We confirmed our hypothesis under various assumptions about the data distribution. Subsequently, we performed Pearson's correlation analysis and Spearman's correlation analysis. Pearson's correlation analysis revealed a linear relationship with normally distributed data. Spearman's correlation analysis relaxed this assumption. This approach can be used when the relationship is monotonic but not necessarily linear. The use of both tests validated our findings under both linear and nonlinear circumstances. This approach enhanced the robustness of our hypothesis testing. These tests confirmed a statistically positive correlation. This correlation existed between the pedestrian's maximum velocity and the potential threat to the pedestrian.
\subsection{Measurement of velocity and density}
Fig. \ref{Fig.2} shows the speed-density fundamental diagram of the different experiments for the different teams based on the pedestrian speed and density calculated from the data. The pedestrian velocity has also been used in other studies.
Based on the velocity calculation in Zhang et al. \citep{zhangtransitions2011}, the velocity of an individual is calculated as follows:

\begin{equation}
  \mathbf{v_i}(t)=\frac{\mathbf{x_i}(t+\Delta t / 2)-\mathbf{x_i}(t-\Delta t / 2)}{\Delta t}
\end{equation}\label{eq_v}
where $\Delta t = $0.64 s corresponds to 16 frames and $\mathbf{x_i}(t)$ refers to the position of pedestrian $i$ at time $t$. A duration of 0.64 seconds is commonly utilized to minimize deviations in the head bobbing. The pedestrian speed in the fundamental diagram is the magnitude of $\mathbf{v}_i(t)$.

We modified the methods used in the study (\citep{parisipedestrian2021}) to compute density features in different motion directions. The density of an individual is calculated as follows:
\begin{equation}
  \rho_i=\frac{k-1}{\pi d_k^2-A_{BS}}
\end{equation}
where $k=5$ indicates that the density zone is a field of view circle with a relative distance to the fifth nearest neighbour (including himself). These fluctuations are all considered within ±60 degrees of the pedestrian's direction of motion. $d_k$ is the radius of the density zone if there are no obstacles within the field of view. The density zone is a circle with a radius of $d_k$. $A_{BS}$ denotes the area within the density circle, but it is not visible to the field of view. The field of view is calculated by the commonly used ray-casting method.

\begin{figure}
  \centering
    \includegraphics{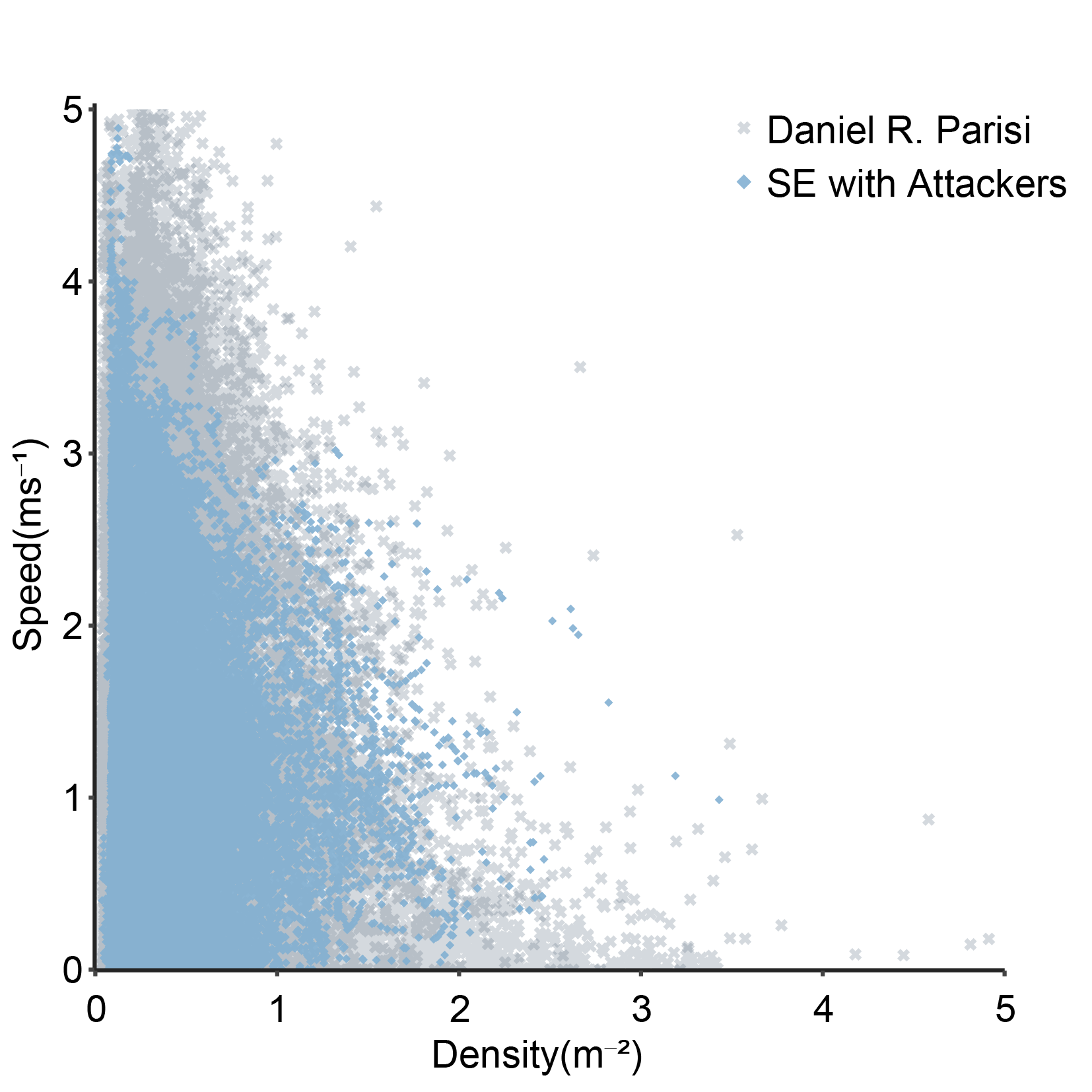}
\caption{\textbf{Comparison of density calculation results.}}\label{Fig.8}
\end{figure}

As shown in Fig. \ref{Fig.8}, we compared the density results calculated by our method with the real-world density results calculated in a previous study (\citep{parisipedestrian2021}) (Dataset3, Dataset5). to confirm the feasibility of the method.

\subsection{Kladek-Newell-Weidmann equation}
The function used to describe the desired velocity curve in Fig. 2 is based on the Kladek-Newell-Weidmann equation outlined in Weidmann's study (\citep{weidmanntransporttechnik1993}). It is calculated as follows:
\begin{equation}
  v(\rho)=v_0(1-e^{-g(\frac{1}{\rho}-\frac{1}{\rho_{max}})})
\end{equation}
where $v_0$ is the desired velocity and $g$ and $\rho_{max}$ are constants. $\rho_{max}$ represents the density at which the speed falls to zero. The parameters of the desired velocity curves in Fig. \ref{Fig.2} are listed in the Appendix, Table \ref{tablea2}. This study used the same value of $g$ as that reported by \citep{parisipedestrian2021} to enable a comparison of these results.

\section*{Acknowledgments}
This work is supported by the National Natural Science Foundation of China (No. 72074149 and No.52072286). The authors thank the anonymous reviewers and the editor for their comments. Particularly, we thank the third reviewer for helping us improve the equation expressions and research content.

\appendix
\renewcommand\thefigure{\Alph{section}\arabic{figure}}
\renewcommand\thetable{\Alph{section}\arabic{table}}
\section{Appendix}
\newpage
\setcounter{figure}{0}
\setcounter{table}{0}
\begin{figure*}[htbp]
\centering
\includegraphics[scale=.9]{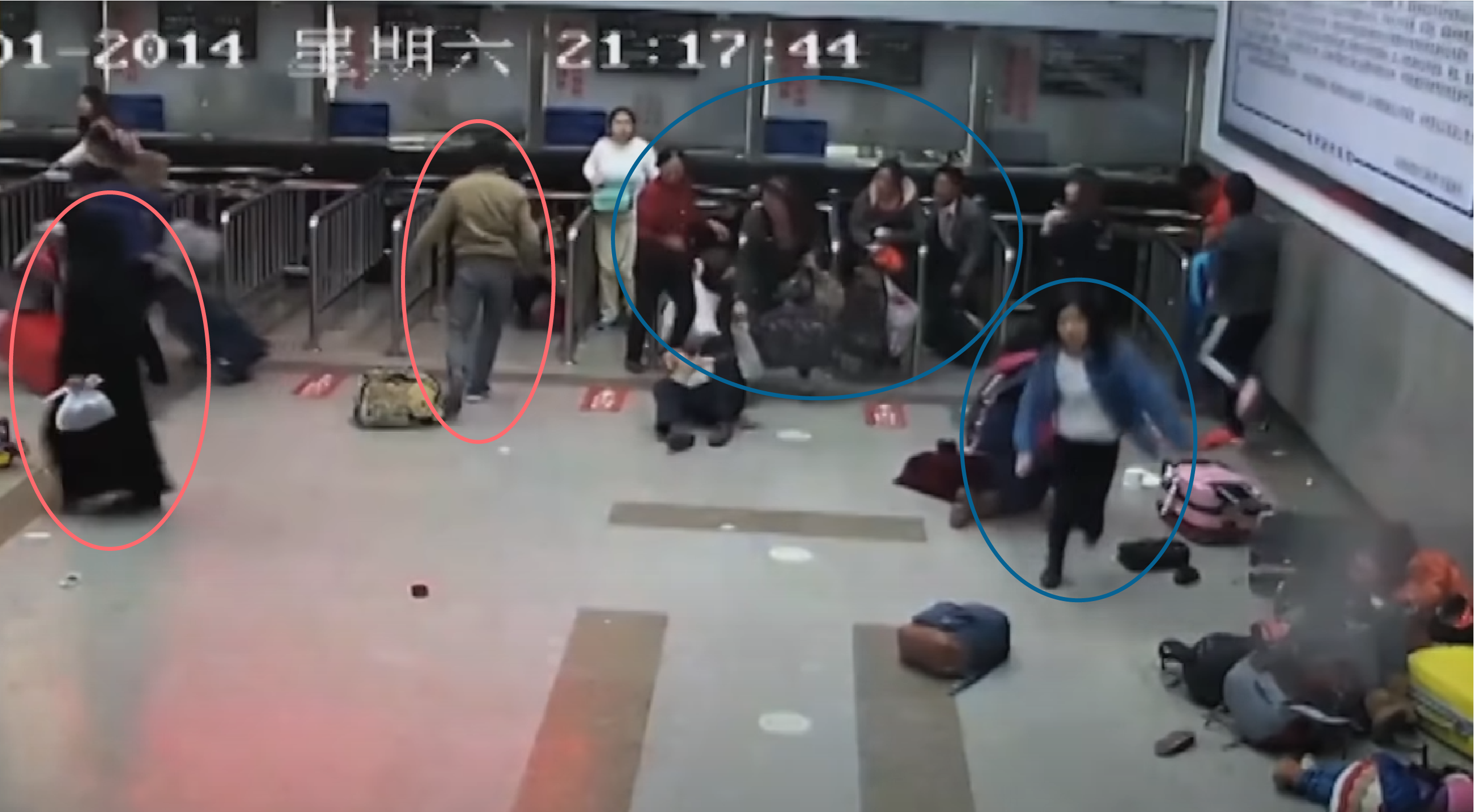}
\caption{\textbf{Screenshot of a real-life example of a pedestrian near the assault target circling around to the rear of the attacker.} A screenshot from a terrorist attack in Kunming, China, in 2014 (\citep{cgtnfighting2019}.) In this image of an attacker launching an attack, the blue circles mark pedestrians close to the assault target, which is directly in front of the attacker. The red circle denotes the attackers.}
\label{figa1}
\end{figure*}

\begin{figure*}[b]
\centering
\includegraphics[scale=.9]{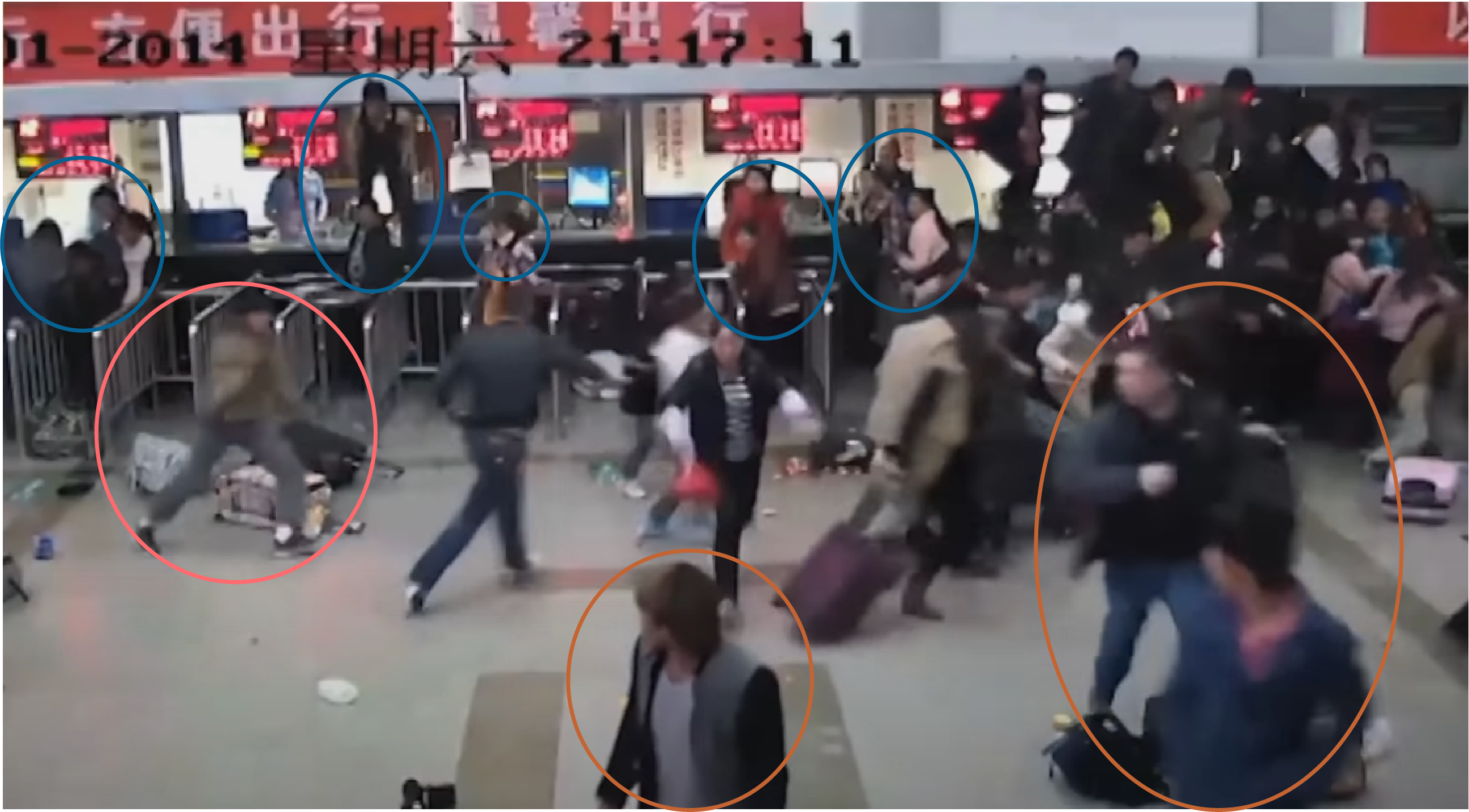}
\caption{\textbf{Screenshot of a real-life case where pedestrians are stationary when attackers block their effective evacuation routes.} A screenshot from a terrorist attack in Kunming, China, in 2014 (\citep{cgtnfighting2019}). The red circle denotes the attacker. The blue circles denote pedestrians whose desired velocity has been halted until the attacker no longer blocks their effective evacuation routes. The pedestrians in orange circles are not directly threatened by the attacker and look back at a low desired velocity.}
\label{figa2}
\end{figure*}

\begin{table*}[h]
\caption{\textbf{Experimental settings.} In the "Exit width" column, the first number indicates the upper exit, and the second number indicates the lower exit. Experimental settings without attackers are only available in the single-exit and exit-obscured chase experiments.}
\label{tablea1}
\begin{tabular}{@{}cccc@{}}
\toprule
Experimental scenes  & Number of Attackers & Number of Pedestrians & Exit width(m)\\
\midrule
Single Exit    & 0   & 50  & 0.8  \\
Single Exit    & 1   & 49  & 0.8  \\
Single Exit    & 2   & 48  & 0.8  \\
Single Exit    & 3   & 47  & 0.8  \\
Dual Exit    & 1   & 49  & 0.8 \& 0.8  \\
Dual Exit    & 2   & 48  & 0.8 \& 0.8  \\
Dual Exit    & 3   & 47  & 0.8 \& 0.8  \\
Exit-Obscured Chase & 0   & 25  & 0.8 \& 0.8  \\
Exit-Obscured Chase & 0   & 25  & 1.2 \& 0.8  \\
Exit-Obscured Chase & 1   & 24  & 0.8 \& 0.8  \\
Exit-Obscured Chase & 1   & 24  & 1.2 \& 0.8  \\
Exit-Obscured Chase & 2   & 23  & 0.8 \& 0.8  \\
Exit-Obscured Chase & 2   & 23  & 1.2 \& 0.8  \\
Exit-Obscured Chase & 3   & 22  & 0.8 \& 0.8  \\
Exit-Obscured Chase & 3   & 22  & 1.2 \& 0.8  \\
Exit-Obscured Chase & 5   & 20  & 0.8 \& 0.8  \\
Exit-Obscured Chase & 5   & 20  & 1.2 \& 0.8  \\
\bottomrule
\end{tabular}
\end{table*}

\begin{table*}[h]
\caption{\textbf{Parameters of the desired curves.} The curves in the table correspond to the curves shown in the respective figure, from top to bottom. Except for the parameter $g$, the velocity curve parameters are mainly determined by the fundamental diagram obtained from the experiments.}
\label{tablea2}
\begin{tabular}{@{}cccc@{}}
\toprule
Desired Velocity Curve  & $v_{0}$ & $\rho_{max}$ & $g$\\
\midrule
Fig. 2 b, c    & 1   & 9  & 9 \\
Fig. 2 b, c    & 2   & 8  & 9 \\
Fig. 2 b, c    & 3   & 5  & 9 \\
Fig. 2 b, c    & 4   & 1.5  & 9 \\
Fig. 2 b, c    & 5   & 0.5  & 9 \\
Fig. 2 e, f    & 1   & 7  & 9 \\
Fig. 2 e, f    & 2   & 5.5  & 9 \\
Fig. 2 e, f    & 3   & 5  & 9 \\
Fig. 2 e, f    & 4   & 1.5  & 9 \\
Fig. 2 e, f    & 5   & 0.5  & 9 \\
\bottomrule
\end{tabular}
\end{table*}
\clearpage
\bibliographystyle{model1-num-names}
% \bibliographystyle{cas-model2-names}

\bibliography{references}

\end{document}